\documentclass[referee]{jfm}

\ifCUPmtlplainloaded \else
  \checkfont{eurm10}
  \iffontfound
    \IfFileExists{upmath.sty}
      {\typeout{^^JFound AMS Euler Roman fonts on the system,
                   using the 'upmath' package.^^J}%
       \usepackage{upmath}}
      {\typeout{^^JFound AMS Euler Roman fonts on the system, but you
                   dont seem to have the}%
       \typeout{'upmath' package installed. JFM.cls can take advantage
                 of these fonts,^^Jif you use 'upmath' package.^^J}%
      }
  \else
  \fi
\fi

\ifCUPmtlplainloaded \else
  \checkfont{msam10}
  \checkfont{msam10}
  \iffontfound
    \IfFileExists{amssymb.sty}
      {\typeout{^^JFound AMS Symbol fonts on the system, using the
                'amssymb' package.^^J}%
       \usepackage{amssymb}%
         \let\leq=\leqslant
       \let\ge=\geqslant  \let\geq=\geqslant
      }{}
  \fi
\fi

\ifCUPmtlplainloaded \else
  \IfFileExists{amsbsy.sty}
    {\typeout{^^JFound the 'amsbsy' package on the system, using it.^^J}%
     \usepackage{amsbsy}}
    {\providecommand\boldsymbol[1]{\mbox{\boldmath $##1$}}}
\fi

\providecommand\bnabla{\boldsymbol{\nabla}}

\newsavebox{\astrutbox}
\sbox{\astrutbox}{\rule[-5pt]{0pt}{20pt}}

\newcommand\etal{\mbox{\textit{et al.}}}

\newcommand\eg{e.g.\ }

\usepackage{times}
\usepackage{epsfig}
\input epsf

\newcommand{\ptl}{\partial}

\title[{Universality of shear-banding instability in granular shear flow}]{Universality of shear-banding instability and  crystallization in sheared granular fluid}
\author[M. Alam, P. Shukla and S. Luding ]%
{
\textsc{MEHEBOOB ALAM$^1$}\footnote{Corresponding author: Email: meheboob@jncasr.ac.in\\
Published in {\it J. Fluid Mech.}, vol. 615, p. 293-321 (2008)},
\textsc{PRIYANKA SHUKLA$^1$}
\textsc{and}
\textsc{STEFAN LUDING$^2$}
}
\affiliation{
$^1$ Engineering Mechanics Unit, Jawaharlal Nehru Center for Advanced
 Scientific Research\\
 Jakkur P.O., Bangalore 560064, India.\\
$^2$MultiScale Mechanics, Faculty of Science and Technology,
University of Twente,
P.O. Box 217, 7500 AE Enschede, The Netherlands
}
\pubyear{2008}
\date{\today}
\begin{document}

\maketitle

\begin{abstract}
The linear stability analysis of an uniform shear flow of  granular
materials is revisited using several cases of a Navier-Stokes'-level constitutive model
in which we incorporate  the global equation of states for pressure and thermal conductivity 
(which are  accurate up-to the maximum packing density $\nu_{m}$) and 
the shear viscosity is allowed to diverge at a  density $\nu_\mu$ ($< \nu_{m}$),
with all other transport coefficients diverging at $\nu_{m}$.
It is shown that the emergence of shear-banding instabilities (for perturbations having no variation
along the streamwise direction), that lead to shear-band
formation along the gradient direction,  depends crucially on the choice of the constitutive model. 
In the framework of a dense constitutive model that incorporates only collisional
transport mechanism, it is shown that an accurate global equation of state for
pressure or a viscosity divergence at a lower density or a stronger viscosity divergence
(with other transport coefficients being given by respective
Enskog values that diverge at $\nu_m$) can induce  shear-banding instabilities, even though
the original dense Enskog model is stable to such shear-banding instabilities.
For any constitutive model, the onset of this shear-banding instability 
is tied to a {\it universal} criterion 
in terms of constitutive relations for viscosity and pressure,
and the sheared granular flow evolves toward a state of lower ``dynamic'' friction,
leading to the shear-induced band formation, as it cannot sustain 
increasing dynamic friction with increasing density to stay in the homogeneous state.
A similar criterion of a lower viscosity or a lower viscous-dissipation is responsible 
for the shear-banding state in many complex fluids.
\end{abstract}

\section{Introduction}
\label{sec1}

One challenge in granular flow research is to devise appropriate
hydrodynamic/continuum models to  describe its macroscopic behavior.
Rapid granular flows (Campbell 1990; Goldhirsch 2003)
can be well modeled by an idealized system of
smooth hard spheres with {\it inelastic} collisions.  The kinetic theory
of dense gases has been modified to  obtain Navier-Stokes-like (NS)
hydrodynamic equations, with an additional equation for the  fluctuation
kinetic energy of particles (i.e., granular energy) that incorporates
the {\it dissipative}-nature of particle collisions. 
Such NS-order hydrodynamic models have  been widely used as  
prototype models to gain insight into the ``microscopic''
understanding of various physical phenomena involved in granular flows.

The plane Couette flow has served as a prototype model problem to study the
rheology (Lun \etal~1984; Jenkins \& Richman 1985; Campbell 1990; Sela \& Goldhirsch 1998;
Alam \& Luding 2003, 2003a; Tsai, Voth \& Gollub 2003; Alam \& Luding 2005; Gayen \& Alam 2008) and 
dynamics (Hopkins \& Louge 1991; McNamara 1993; Tan \& Goldhirsch 1997;
Alam \& Nott 1997, 1998;  Conway \& Glasser 2004;
Alam \etal~2005; Alam 2005, 2006; Gayen \& Alam 2006;
Saitoh \& Hayakawa 2007) of granular materials.
In the rapid shear flow, the linear stability analyses of plane 
Couette flow (Alam \& Nott 1998;  Alam 2006) showed
that this flow admits different types of stationary and travelling-wave instabilities,
leading to pattern formation. One such instability is the ``shear-banding'' instability
in which the homogeneous shear flow breaks into alternating dense
and dilute regions of particles along the gradient direction. 
This is dubbed ``shear-banding'' instability since the ``nonlinear'' saturation
of this instability (Alam \& Nott 1998;  Alam \etal~2005;
Shukla \& Alam 2008) leads to alternate layers of dense and dilute
particle-bands in which the shear-rate is high/low in
dilute/dense regions, respectively, leading to ``shear-localization'' (Varnik \etal~2003).
This is  reminiscent of shear-band formation in shear-cell experiments
(Savage \& Sayed 1984; Losert \etal~2000; Mueth \etal~2000; Alam \& Luding 2003;
Tsai, Voth \& Gollub 2003):
{\it when a dense granular material is sheared
the shearing is confined within a few particle-layers (i.e., a shear-band)
and  the rest of the material remains unsheared}, 
leading to the two-phase flows of dense and dilute regions.
Previous works (Alam \& Nott 1998;  Alam \etal~2005) showed that the kinetic-theory-based 
hydrodynamic models are able to predict the co-existence of dilute and dense regimes
of such shear-banding patterns.

The above problem has recently been reanalyzed (Khain \& Meerson 2006),
with reference to shear-band formation, with a constitutive model which is 
likely to be valid  in the dense limit.
These authors showed that  their `dense' constitutive model
does not admit shear-banding instabilities of Alam \& Nott (1998);
however, a single modification that the shear
viscosity diverges at a  density lower than other transport coefficients
resulted in the appearance of two-phase-type solutions of dilute and dense flows
that are reminiscent of shear-banding instabilities.
That the viscosity diverges stronger/faster than other transport coefficient has also been
incorporated previously in a constitutive model by Losert \etal~(2000)
that yields a satisfactory  prediction for shear-bands 
in an experimental Couette flow in three-dimensions.

We revisit this problem  to understand the influence of different 
Navier-Stokes'-order constitutive models 
(as detailed in \S2) on the shear-banding instabilities in granular plane Couette flow.
More specifically, we will pinpoint how the effects of  various models,
some of them involving the global equation of state 
and the viscosity divergence (at a density  lower than the maximum packing),
change the shear-banding instabilities predicted by Alam \& Nott (1998).
For the sake of a systematic overview, we will also discuss 
about the instability results based on special limiting cases of these models.
One important finding is that a global equation of state ({\it without viscosity divergence
at a lower density})  leads to a shear-banding instability in the
framework of a ``dense'' model that incorporates only collisional transport mechanism.
Even with the local equation of state, if we use a constitutive relation for viscosity
that  has a stronger divergence (at the same maximum packing density)
than other transport coefficients, we recover shear-banding instabilities in the
framework of the dense-model of Haff (1983).
This brings us to a  {\it crossroad}: is there any connection among
the results of Alam \& Nott, Khain \& Meerson, and the present work?
Is there any {\it universal} criterion for the onset of 
the shear-banding instability in granular shear flow?
Such an universality for the shear-banding instability
indeed exists, solely in terms of the constitutive relations, as we show in this paper.
Possible connections of the present criterion 
of a lower dynamic friction for the shear-banding state 
to explain the onset of shear-banding in many complex fluids
as well as in an elastic hard-sphere fluid are discussed.

\section{Balance equations and constitutive model}

We use a Navier-Stokes-level hydrodynamic model for which we need 
balance equations for mass, momentum and granular temperature:
\begin{eqnarray}
   \left(\frac{\ptl}{\ptl t} + {\bf u}\cdot{\bnabla}\right)\varrho&=& - \varrho\bnabla\cdot{\bf u} \\
   \varrho\left(\frac{\ptl}{\ptl t} + {\bf u}\cdot{\bnabla}\right){\bf u} &=& - \bnabla\cdot{\bf P} \\
   \frac{\dim}{2}\varrho\left(\frac{\ptl}{\ptl t} + {\bf u}\cdot{\bnabla}\right)T &=& 
       - \bnabla\cdot{\bf q} - {\bf P}:\bnabla{\bf u} - {\mathcal D} .
\end{eqnarray}    
Here $\varrho=mn=\rho_p\nu$ is the mass-density, $m$ the particle mass,
$n$ the number density, $\rho_p$ the
material density and $\nu$ the area/volume fraction of particles;
${\bf u}$ is the coarse-grained velocity-field and $T$ is the granular temperature of the fluid.
Note that the granular temperature, $T =\langle C^2/\dim\rangle$, is defined as the mean-square
fluctuation velocity, with ${\bf C}=({\bf c} -{\bf u})$ being the peculiar velocity of particles
and ${\bf c}$ the instantaneous particle velocity; $\dim$ is the dimensionality
of the system and here onward we focus on the two-dimensional ($\dim=2$) system of an
inelastic ``hard-disk'' fluid.
The flux terms are  the stress tensor, ${\bf P}$, and the granular heat flux, ${\bf q}$;
${\mathcal D}$ is the rate of dissipation of granular energy per unit volume--  for these three terms
we need appropriate constitutive relations which are detailed below.

\subsection{General form of Newtonian constitutive model:  Model-$A$}

The standard Newtonian form of the stress tensor and the Fourier law of heat flux 
are:
\begin{eqnarray}
   {\bf P} &=& (p - \zeta\bnabla\cdot{\bf u}){\bf I} - 2\mu {\bf S}, \\
   {\bf q} &=& -\kappa \bnabla T,
\end{eqnarray}
where ${\bf I}$ is the identity tensor and
${\bf S}$ the deviator of the deformation rate tensor.
Here $p$, $\mu$, $\zeta$ and $\kappa$ are  pressure, shear viscosity, bulk viscosity
and thermal conductivity of the granular fluid, respectively.

Focussing on the nearly elastic limit ($e\to 1$) of an inelastic hard-disk 
(of diameter $d$) fluid,
the constitutive expressions for $p$, $\mu$, $\zeta$, $\kappa$ and $\mathcal D$
are given by 
\begin{equation}
\begin{array}{rcl}
     p(\nu,T) &=& \rho_pf_1(\nu)T, \quad\quad
     \mu(\nu,T) \;=\; \rho_p d f_2(\nu)\sqrt{T},\\
    \zeta(\nu,T) &=& \rho_p d f_3(\nu)\sqrt{T}, \quad
    \kappa(\nu,T) \;=\; \rho_p d f_4(\nu)\sqrt{T},\\
    {\mathcal D}(\nu,T) &=& \frac{\rho_p}{ d} f_5(\nu,e)T^{3/2},
\end{array}
\end{equation}
where $f_1$--$f_5$ are non-dimensional functions of the particle area fraction $\nu$
(Gass 1971; Jenkins \& Richman 1985):
\begin{eqnarray}
   f_1(\nu) &=&  \nu + 2\nu^2\chi,\\
   f_2(\nu) &=& \frac{\sqrt{\pi}}{8\chi} + \frac{\sqrt\pi}{4}\nu 
                + \frac{\sqrt\pi}{8}\left(1 + \frac{8}{\pi}\right)\nu^2\chi, \\
  f_3(\nu) &=& \frac{2}{\sqrt\pi}\nu^2\chi,\\
  f_4(\nu) &=& \frac{\sqrt\pi}{2\chi} + \frac{3\sqrt\pi}{2}\nu 
      + \sqrt\pi \left(\frac{2}{\pi} +\frac{9}{8}\right)\nu^2\chi, \\
  f_5(\nu,e) &=& \frac{4}{\sqrt{\pi}}(1-e^2)\nu^2\chi .
\end{eqnarray}
These constitutive expressions give good predictions
for transport coefficients of nearly elastic granular fluid
up-to a density of $\nu\approx 0.55$ (see, figure 2 of Alam \& Luding 2003a).
In the above expressions, $\chi(\nu)$ is the contact radial distribution function 
which is taken to be of the following form (Henderson 1975)
\begin{equation}
     \chi(\nu) = \frac{1- 7\nu/16}{(1- \nu/\nu_m)^2},
\label{eqn_rdf1}
\end{equation}
that diverges at some finite density $\nu=\nu_m$.
For the ideal case of point particles
(i.e. in one dimension, Torquato 1995), we have $\nu_m=1$ which is
unrealistic for macroscopic grains at very high densities;  
there are two other choices for this diverging density:
the random close packing density $\nu_r=\nu_m=0.82$ 
or the maximum packing density $\nu_m=\pi/2\sqrt{3}\approx 0.906$ in two-dimensions
(Torquato 1995); up-to some moderate density ($\nu\sim 0.5$), there is
no difference in the value of $\chi(\nu)$ for any choice of $\nu_m=1$ or $0.906$ or $0.82$.
The range of validity of different variants of model radial distribution functions
is discussed in an upcoming paper (Luding 2008).

We shall denote the above constitutive model (2.6--2.11), with
the contact radial distribution function being given by (2.12), as ``model-$A$''.  
Since the stability results do not differ qualitatively 
with either choice of the numerical value for $\nu_m$ ($=0.82$ or $0.906$ or $1$),
we will present all results with  $\nu_m=\pi/2\sqrt{3}$ in (2.12)
(except in figure 12$d$, see \S5.2).

Now we consider the dilute and dense limits of model-$A$.
It should be noted that each transport coefficient has contributions from the
`kinetic' and `collisional' modes of transport: while the former is dominant
in the Boltzmann limit ($\nu\to 0$), the latter is dominant in the dense limit ($\nu\to \nu_m$).    
For example, the pressure can be decomposed into its kinetic and collisional contributions:
\begin{equation}
     p = p^k + p^c = \rho_p (f_1^k + f_1^c) T .
\end{equation}
To obtain the constitutive expressions for the
dilute and dense regimes, one has to  take the appropriate limit of
all functions  $f_1$--$f_4$:
\begin{equation}
\begin{array}{rcl}
   f_1 &=& f_1^k + f_1^c , \quad
   f_2 \;=\; f_2^k + f_2^c, \\
   f_3 &=& f_3^k + f_3^c , \quad
   f_4 \;=\; f_4^k + f_4^c .
\end{array}
\end{equation}
Based on this decomposition, we have the following two limiting cases of model-$A$.

\subsubsection{Dilute limit: Model $A_0$}

For this dilute ($\nu\to 0$) model, $\chi(\nu)\to 1$ and
the constitutive expressions are assumed to contain
contributions only from the kinetic mode of transport:
\begin{equation}
\begin{array}{rcl}
   f_1 \equiv f_1^k(\nu) &=&  \nu, \quad\quad
   f_2 \equiv f_2^k(\nu)  \;=\; \frac{\sqrt{\pi}}{8} + \frac{\sqrt\pi}{4}\nu, \\
   f_3 \equiv f_3^k(\nu) &=& 0, \quad\quad
   f_4 \equiv f_4^k(\nu) \;=\; \frac{\sqrt\pi}{2} + \frac{3\sqrt\pi}{2}\nu, \\
   f_5 \equiv f_5(\nu\to 0) &=& \frac{4}{\sqrt{\pi}}(1-e^2)\nu^2 .
\end{array}
\end{equation}
We shall call this ``model $A_0$''.
Note that $f_5$ has only the leading-order collisional contribution, since no energy is dissipated
in kinetic (free-flow) motion.

\subsubsection{Dense  limit: Model $A_d$}

For this  {\it dense} model, the constitutive expressions are assumed to contain
contributions only from collisional mode of transport:
\begin{equation}
\begin{array}{rcl}
    f_1 \equiv f_1^c(\nu) &=&   2\nu^2\chi, \quad\quad\quad
    f_2 \equiv f_2^c(\nu) \; =\; 
                   \frac{\sqrt\pi}{8}\left(1 + \frac{8}{\pi}\right)\nu^2\chi, \\
    f_3 \equiv f_3^c(\nu) &=& \frac{2}{\sqrt\pi}\nu^2\chi,  \quad\quad
    f_4 \equiv f_4^c(\nu) \;=\; 
          {\sqrt\pi}\left(\frac{2}{\pi} +\frac{9}{8}\right)\nu^2\chi,\\ 
    f_5 \equiv f_5(\nu,e) &=& \frac{4}{\sqrt{\pi}}(1-e^2)\nu^2\chi .
\end{array}
\end{equation}
We shall call this ``model $A_d$'' which  is nothing but Haff's model (1983).

\subsection{Model $B$: global equation of state (EOS)}

In two-dimensions, a global equation of state for pressure has recently
been proposed by Luding (2001):
\begin{equation}
    p/\rho_p T \equiv f_1(\nu) = \nu  + \nu\left( P_4 + m(\nu)\left[P_{dense}-P_4\right]\right) ,
    \quad \forall \quad 0\leq \nu\leq \nu_m = \frac{\pi}{2\sqrt{3}},
    \label{eqn_GEOP1}
\end{equation}
where
\begin{equation}
\begin{array}{rcl}
   P_4 &=& 2\nu\chi_{4},\\
   \chi_{4}(\nu) &=& \frac{1-7\nu/16}{(1-\nu)^2}  - \frac{\nu^3}{128(1-\nu)^4},\\
   P_{dense} &=& \frac{2\nu_m h_3(\nu_m-\nu)}{(\nu_m-\nu)} -1, \\
   h_3(\nu_m -\nu) &=& 1 -0.04(\nu_m -\nu) + 3.25 (\nu_m -\nu)^3,   \\
   m(\nu) &=& \left[1 + \exp(-(\nu-\nu_f)/m_0)\right]^{-1} .
\end{array}
\label{eqn_GEOP2}
\end{equation}
Here $\nu_f$ is the freezing point density, and $m(\nu)$ is a merging function 
that selects $P_4$ for $\nu<<\nu_f$ and $P_{dense}$ for $\nu>>\nu_f$;
the value of $m_0$ is taken to be $0.012$ along with a freezing density of $\nu_f=0.7$.
It should be noted that the above functional form of $f_1(\nu)$,
(2.17), is a monotonically increasing function of $\nu$,
and it has been verified (Luding 2001, 2002; Garcia-Rojo, Luding \& Brey 2006)
from molecular dynamics simulations of elastic hard-disk systems 
that the numerical values for different constants in (\ref{eqn_GEOP2}) are
accurate (within much less than $1\%$ except around $\nu_f$) up-to the maximum packing density.

Rewriting  equation~(\ref{eqn_GEOP1}) as
\begin{equation}
   f_1(\nu) = \nu + 2\nu^2\chi^p(\nu),
\end{equation}
we can define an ``effective'' contact radial distribution function for pressure:
\begin{eqnarray}
  \chi^p(\nu) &=& \frac{1}{2\nu}\left( P_4 + m(\nu)\left[P_{dense}-P_4\right]\right), \nonumber \\
              &=& \chi_{4}  + m(\nu)\left[
                     \frac{ h_3(\nu_m-\nu)}{\nu(1-\nu/\nu_m)} - \frac{1}{2\nu}  -  \chi_{4}
                                      \right], 
\quad \mbox{with} \quad \nu_m =\frac{\pi}{2\sqrt{3}}.
\end{eqnarray} 
It should be noted here that this functional form of $\chi^p(\nu)$ has been verified
by molecular dynamics simulations of plane shear flow of frictional inelastic disks
(Volfson, Tsimring \& Aranson 2003). In particular, they found that
the simulation data on $G(\nu)=\nu\chi^p(\nu)$ agree with the predictions
of (2.20), however, its divergence appears to occur at the random close packing limit 
in two-dimensions ($\nu_m=\nu_r=0.82$).

Similar to pressure, a global equation for thermal conductivity has been suggested
by Garcia-Rojo \etal~(2006) which is also accurate up-to the maximum packing density.
More specifically, the non-dimensional function of density  for thermal conductivity 
(equation~(2.10)) is replaced by
\begin{equation}
    f_4(\nu) = \frac{\sqrt\pi}{2\chi^\kappa(\nu)} + \frac{3\sqrt\pi}{2}\nu 
                  + \sqrt\pi \left(\frac{2}{\pi} +\frac{9}{8}\right)\nu^2\chi^{\kappa}(\nu) \\
\end{equation}
where $\chi^\kappa(\nu)$ is the ``effective'' 
contact radial distribution function for thermal conductivity:
\begin{equation}
  \chi^{\kappa}(\nu)= 
       \chi(\nu,\nu_m=1)  + m(\nu)\left[
        \frac{h_3(\nu_m-\nu)}{\nu(1-\nu/\nu_m)} - \frac{1}{2\nu} - \chi(\nu,\nu_m=1) \right] ,
\end{equation}
with $\chi(\nu,\nu_m=1)$ being obtained from (2.12) by putting $\nu_m=1$.

The model-$A$ with the above global equation of state for pressure and 
the global equation for thermal conductivity is called  ``model-$B$''.
Note that the dilute limit of model-$B$  is the same as that of model-$A$,
however, the dense limits of both models are different in the choice
of the equation of state and thermal conductivity.

Now we identify two subsets of model-$B$:  ``model-$B^p$'' and ``model-$B^{\kappa}$'' where 
the former is model-$B$ with a global equation of state for pressure (2.17) only
and the latter is model-$B$ with a global equation for thermal conductivity (2.21) only.
As clarified in the previous paragraph, the 
remaining transport coefficients of model-$B$ are same as in model-$A$.

\subsection{Model-$C$:  viscosity divergence}

Here we consider a variant of model-$A$ which incorporates another
ingredient in the constitutive model: {\it the shear viscosity
diverges at a density lower than other transport coefficients} (Garcia-Rojo \etal~2006).
This can be incorporated in the corresponding dimensionless function  
for the shear viscosity (2.8):
\begin{equation}
   f_2^\mu(\nu) = f_2(\nu)\left( 1 + \frac{0.037}{\nu_\mu - \nu}\right), 
\end{equation}
which diverges at a density, $\nu=\nu_\mu <\nu_m$, that is lower than the
close packing density.  
The first term on the right-hand-side is the standard Enskog term (2.8) and the
second term incorporates a correction due to the viscosity divergence.
For all results shown here, we use $\nu_\mu=0.71$
which was observed  for the unsheared case (Garcia-Rojo \etal~2006); note, however, that the precise
density at which viscosity diverges in a shear flow can be larger than $0.71$
(see figure~6 of Alam \& Luding 2003).

The model with all transport coefficients as in model-$A$ but with its
viscosity divergence being at a lower density is termed as  ``model-$C$''.
Similar to  model-$A$, we can recover the dilute and dense limits of model-$C$,
by  separating the kinetic  and collisional contributions to each transport coefficient.
The dense limit of model-$C$ is, however, reached at $\nu=\nu_\mu$
due to the viscosity divergence.

\subsection{Model-$D$:  Global equation of state and viscosity divergence}

This is the most general model in which we incorporate:
(1) the global equation of state for pressure (2.17),
(2) the global equation for thermal conductivity (2.21), and 
(3) the viscosity divergence (2.23).
The other transport coefficients of model-$D$ are same as in model-$A$.

\section{Plane shear and linear stability}

Let us consider the plane shear flow of granular materials between two walls
that are separated by a distance $\tilde{H}$:
the top wall is moving to the right with a velocity $U_w/2$ and the bottom wall is 
moving to the left with the same velocity.
We impose no-slip and zero heat-flux boundary conditions at both walls:
\begin{equation}
   {\bf u}= \pm U_w/2
   \quad   \mbox{and}   \quad
   {\bf q} =0
   \quad   \mbox{at}   \quad
   y = \pm \tilde{H}/2.
\label{eqn_BC1}
\end{equation}

The equations of motion for the steady and fully developed shear flow admit
the following solution:
\begin{equation}
   {\bf u} = [u(y), v(y)] = [(U_w/\tilde{H})y, 0],
   \quad
   \nu(y)=const. =\overline{\nu},
   \quad
   T = d^2\left(\frac{{\rm d}u}{{\rm d}y}\right)^2\frac{f_2(\nu)}{f_5(\nu)}.
 \label{eqn_usf1}
\end{equation}
The shear rate, ${\rm d}u/{\rm d}y = U_w/\tilde{H}$, is 
uniform (constant) across the Couette gap,
and this solution will henceforth be  called  {\it uniform shear solution}.
Note that if viscosity diverges at $\nu=\nu_\mu$, there is {\it no} uniform
shear solution for $\nu > \nu_\mu$,
i.e. the uniform shear flow is possible only for densities $0<\nu<\nu_\mu$.

We have non-dimensionalized all quantities by using
$\tilde{H}$, $\tilde{H}/U_w$ and $U_w$ as the reference
length, time and velocity scales, respectively.
The explicit forms of dimensionless balance equations
as well as the dimensionless transport coefficients are written down in Appendix A.
There are three dimensionless control parameters to characterize
our problem: the scaled Couette gap $H=\tilde{H}/d$,
the mean density (area fraction) $\nu=\overline\nu$
and the restitution coefficient $e$.
Here onwards, all quantities are expressed in dimensionless form.

\subsection{Linear stability}

The  stability analysis of the plane shear flow has been thoroughly 
investigated (Alam \& Nott 1998; Alam 2005, 2006; Alam \etal~2005) using the 
constitutive models of class $A$ for which all transport coefficients diverge at 
the maximum packing fraction $\nu\to\nu_m$, as outlined in \S2.1.
The same analysis is carried out here for a specific type of perturbations 
that are {\it invariant} along the streamwise/flow ($x$) direction, 
having variations  along the gradient ($y$) direction only. This implies
that the $x$-derivatives of all quantities are set to zero 
($\partial/\partial x (\cdot) =0$) in the governing equations.
The analysis being identical with that of Alam and Nott (1998), we refer the
readers to that article for mathematical details.

Consider the stability of the uniform shear solution (\ref{eqn_usf1}) against
perturbations that have spatial variations 
along the $y$-direction only, \eg the density field can be written as
$\nu(y,t) = \nu + \nu'(y,t)$,
with the assumption of small-amplitude perturbations, $|\nu'(y,t)/\nu| <<1$,
for the linear analysis.
Linearizing around the uniform shear solution, we obtain a set of
partial differential equations:
\begin{equation}
    \frac{\ptl X}{\ptl t} = {\mathcal L} X,
      \quad
      \mbox{with}
      \quad
   {\mathcal B}X\equiv \left(1, 1, \frac{\rm d}{{\rm d}y}\right)\cdot(u',v',T')=0,
\end{equation}
where $X=(\nu', u', v', T')$ is the vector of perturbation fields,
$\mathcal L$ is the linear stability operator
and ${\mathcal B}$ denotes the boundary operator (i.e.,
zero slip, zero penetration and zero heat flux).
Assuming exponential solutions in time $X(y,t)=\hat{X}(y)\exp(\omega t)$,
we obtain a  differential-eigenvalue problem:
\begin{equation}
        \omega \hat{X} = {\mathcal L}\left(\frac{{\rm d}^2}{{\rm d} y^2}, 
   \frac{\rm d}{{\rm d} y},...\right)\hat{X}, 
      \quad
      \mbox{with}
      \quad
   {\mathcal B}\hat{X}=0,
  \label{eqn_eig1}      
\end{equation}
where $\hat{X}(y) =(\hat{\nu}, \hat{u}, \hat{v}, \hat{T})(y)$ is the
unknown disturbance vector that depends on $y$.
Here, $\omega=\omega_r + i\omega_i$ is the complex frequency
whose real part $\omega_r$ denotes the growth/decay rate of perturbations
and the imaginary part $\omega_i$ is its frequency which 
characterizes the propagating ($\omega_i\neq 0$) or stationary ($\omega_i= 0$)
nature of the disturbance.
The flow is stable or unstable if $\omega_r<0$ or $\omega_r>0$, respectively.

\subsection{Analytical solution: dispersion relation and its roots}

Before presenting numerical stability results (\S4), we recall that
the above set of ordinary differential equations (\ref{eqn_eig1})
admits an analytical solution (Alam \& Nott 1998):
\begin{eqnarray}
   ( \hat {\nu}(y), \hat{T}(y)) &=& (\nu_1, T_1)\cos k_n(y\pm 1/2) \\
   ( \hat {u}(y), \hat{v}(y)) &=& (u_1, v_1)\sin k_n(y\pm 1/2),
\end{eqnarray}
where $k_n=n\pi$ is the `discrete' wavenumber along $y$, with $n=1,2,\ldots$ being 
the mode number that tells us the number of zero-crossing  of the density/temperature
eigenfunctions along $y\in(-1/2,1/2)$.
With this, equation~(\ref{eqn_eig1}) boils down
to an algebraic eigenvalue problem:
\begin{equation}
       {\mathcal A}X_1 = \omega X_1,
\end{equation}
where $X_1 = (\nu_1, u_1, v_1, T_1)$   and   ${\mathcal A}$ is a  $4\times 4$ matrix.
This  leads to a fourth-order dispersion relation in $\omega$:
\begin{equation}
     \omega^4 + a_3 \omega^3 + a_2\omega^2 + a_1\omega + a_0 =0,
\label{eqn_dispersion1}
\end{equation}
with coefficients 
\begin{equation}
\begin{array}{rcllcl}
   a_0  &=&  \frac{1}{H^4} a_{04} + \frac{1}{H^6} a_{06},
   & a_1  &=&  \frac{1}{H^2} a_{12} + \frac{1}{H^4} a_{14} + \frac{1}{H^6} a_{16}, \\
   a_2  &=&  \frac{1}{H^2} a_{22} + \frac{1}{H^4} a_{24},  
   & a_3  &=& a_{30} + \frac{1}{H^2} a_{32}. 
\end{array}
\end{equation}
Here $a_{ij}$'s are real functions of density ($\nu$), temperature ($T$), 
and the restitution coefficient ($e$) whose explicit forms  are written down in Appendix B.

Out of four roots of (\ref{eqn_dispersion1}), two roots are real and the other two form
a complex-conjugate pair. It is possible to obtain an approximate
analytical solution for these four roots using the  standard asymptotic
expansion for {\it large} Couette gaps ($H\equiv \tilde{H}/d$), 
with the corresponding small parameter being $H^{-1}$.
The real roots have the  approximations for large $H$:
\begin{eqnarray}
 \omega^{(1)} &=& - \frac{1}{H^2}\left(\frac{a_{04}}{a_{12}}\right)
                      + O\left(H^{-4}\right), \label{eqn_root1} \\
 \omega^{(2)} &=&  \omega_0^{(2)} + \frac{1}{H^2}\frac{\left[a_{12} + a_{22}\omega_0^{(2)}
                       + a_{32}{\omega_0^{(2)}}^2\right]}{\omega_0^{(2)}\left(3a_{30}
                       + 4\omega_0^{(2)}\right)} + O\left(H^{-4}\right),  \label{eqn_root2} 
\end{eqnarray}
where
\begin{equation}
  \omega_0^{(2)} = -\frac{1}{\nu} f_5{T}^{1/2} < 0.
\end{equation}
The real and imaginary parts of the conjugate pair,
\begin{equation}
 \omega^{(3,4)} = \omega_r^{(3,4)} \pm {\rm i}\; \omega_i^{(3,4)},
\label{eqn_TW}
\end{equation}
have the asymptotic approximations for large $H$:
\begin{eqnarray}
{\omega_r}^{(3,4)}
              &=&  { \frac{1}{H^2}\left[\frac{a_{04} + \left(\frac{a_{12}^2}{a_{30}^2}\right)
                      - \left(\frac{a_{12}a_{22}}{a_{30}}\right) }
                {2a_{12}}\right] + O\left(H^{-4}\right) },  \label{eqn_root34real}\\
(\omega_i^{(3,4)})^2
               &=& \frac{1}{H^2}\left(\frac{a_{12}}{a_{30}}\right)
                    + O\left(H^{-4}\right) . \label{eqn_root34img}  
\end{eqnarray}
For  full models (i.e., $A$, $B$, $C$ and $D$), 
it can  be verified that $\omega_r^{(3,4)}$ is always negative, making the first
real root $\omega^{(1)}$, given by (\ref{eqn_root1}), the least-stable mode. 
However, for the dense models (i.e., $A_d$, $B_d$, $C_d$ and $D_d$),
$\omega_r^{(3,4)}$ could be positive, making the travelling waves,  given by (\ref{eqn_TW}),
the least-stable mode at low densities.
These predictions  have been verified 
against numerical values obtained from spectral method as discussed in \S4.

\section{Stability results: comparison among different  models}

For results in this section,
the differential eigenvalue problem (\ref{eqn_eig1}) has been discretized 
using the Chebyshev spectral method
(Alam \& Nott 1998) and the resultant algebraic eigenvalue problem
has been solved using the QR-algorithm of the Matlab-software.
The degree of the Chebyshev polynomial was set to $20$
which was found to yield accurate eigenvalues. 
In principle, we could solve (3.8) to obtain eigenvalues,
but it provides eigenvalues only for a given mode number $n=1,2,\ldots$
in one shot; therefore, one has to solve (3.8) for several $n$ to determine
the growth rate of the most unstable mode.
The advantage of the numerical solution of (\ref{eqn_eig1})
is that it provides all leading eigenvalues in one shot.

\subsection{Results for model-A and its dilute and dense limits}

As mentioned before, the stability analysis of the uniform shear flow with a 3D-variant 
(i.e. for spheres) of model-$A$ 
has been performed before (Alam \& Nott 1998; Alam 2005, 2006; Alam \etal~2005).
Even though the results for our 2D-model are similar
to those for the 3D-model, we show a few representative
results for this constitutive model for the sake of 
a complete, systematic study and for comparison with other models;
note that the results for model-$A_0$ and model-$A_d$ are new.

\begin{figure}
\includegraphics[width=8.0cm]{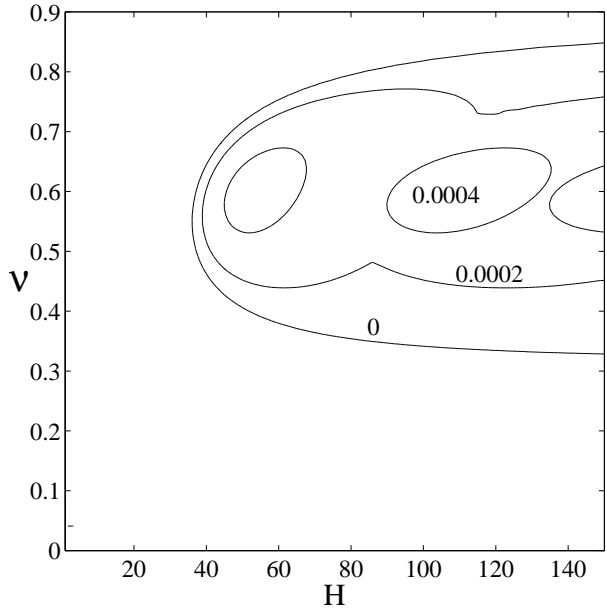}
\caption{
Phase diagram, showing the positive growth-rate contours, for  model-A: 
$e=0.9$, $\nu_{m}=\pi/2\sqrt{3}$.
The flow is unstable inside the neutral contour, denoted by '$0$', and stable outside.
}
\label{fig:fig1}
\end{figure}

The phase diagram, separating the zones of stability and instability,
in the $(\nu, H)$-plane is shown in figure~1 for model-$A$.  
The flow is {\it unstable} inside the neutral contour, denoted by `$0$', and {\it stable} outside;
a few positive growth-rate contours are also displayed.
For the same parameter set, from the respective contours of the frequency, $\omega_i$, 
in the $(\nu, H)$-plane
it has been verified that these instabilities are {\it stationary}, i.e., $\omega_i=0$. 
It is seen that there is a minimum value of the Couette gap ($H=H_{cr}$)  and a minimum
density ($\nu=\nu_{cr}$) below which the flow remains  stable.
With increasing value of $e$, the neutral contour shifts towards the right, 
i.e., $H_{cr}$ increases and hence the flow becomes more stable with increasing $e$.
We shall discuss the dependence of $e$ on the shear-banding instability
and the related instability length scale in \S5.2.

\begin{figure}
\includegraphics[width=7.0cm]{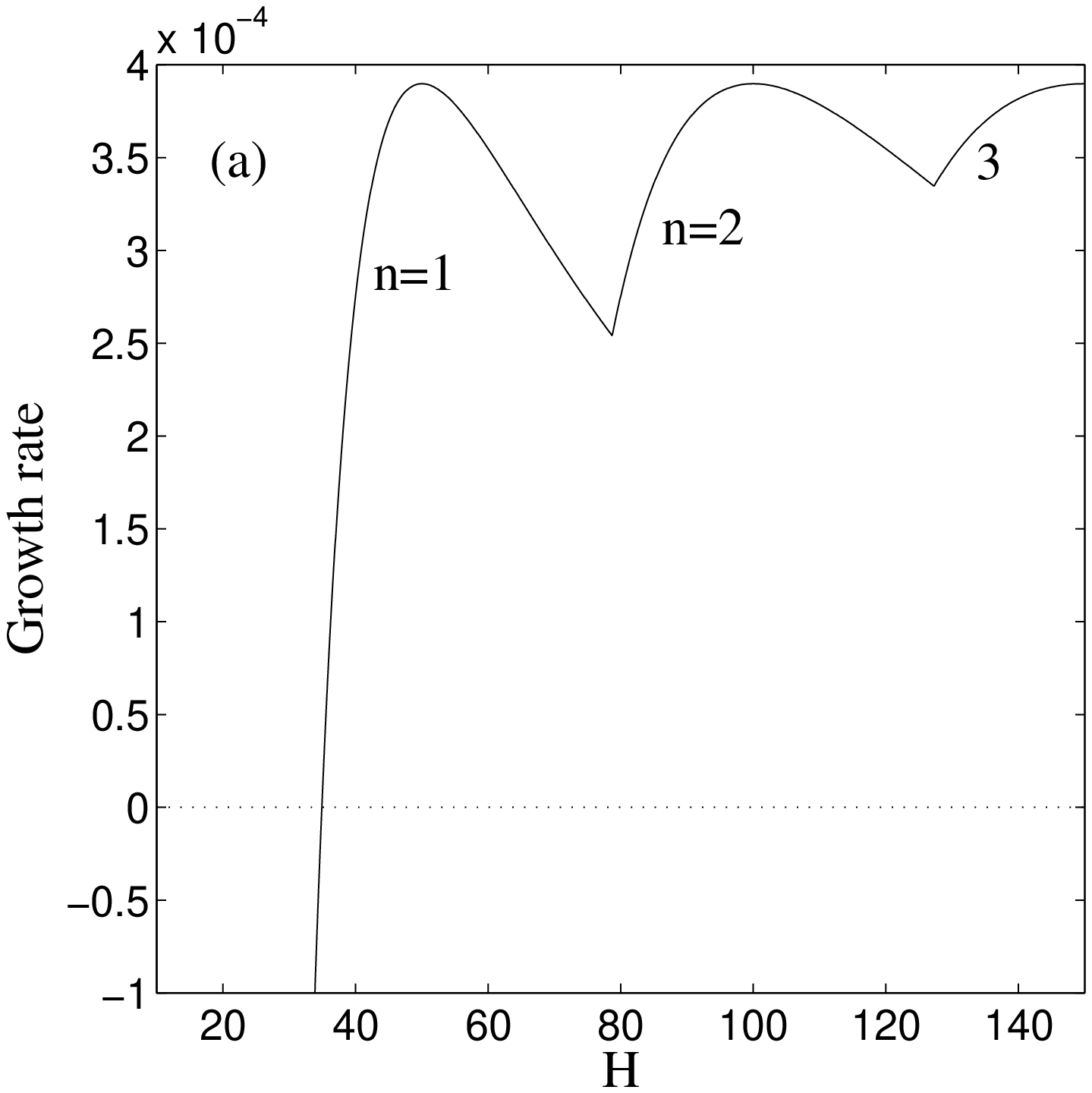}
\includegraphics[width=7.0cm]{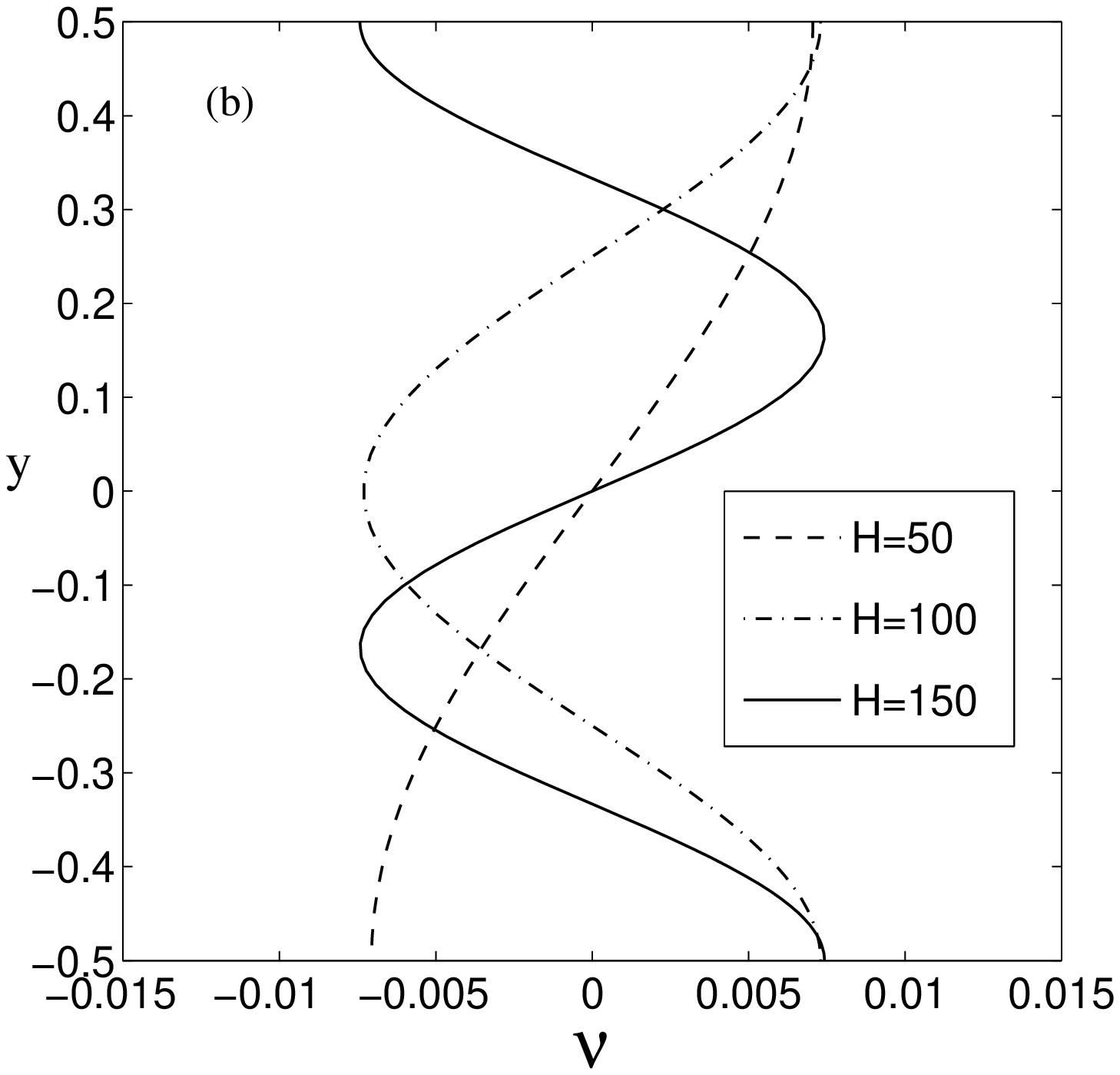}
\caption{
($a$) Variation of the growth rate of the least stable mode with $H$ for $\nu=0.5$ and $e=0.9$.
($b$) Density eigenfunctions, $\hat{\nu}(y)$, of the least stable mode at $H=50$, $100$ and $150$.
}
\label{fig:fig2}
\end{figure}

Figure~\ref{fig:fig2}($a$) shows the variation of the growth rate of the 
least stable mode, $\omega_r^l =\max \omega_r$, with Couette gap for $\nu=0.5$,
with other parameters as in figure~\ref{fig:fig1}. The kinks on the growth-rate
curve correspond to crossing of modes $n=1,2,3,\cdots$. This can be verified
from figure~\ref{fig:fig2}($b$) which displays density eigenfunctions
for three values of Couette gaps $H=50$, $100$ and $150$.
The density eigenfunction at $H=50$ corresponds
to the mode $n=1$ (i.e. $\hat\nu\sim \cos(\pi(y\pm 1/2))=\sin(\pi y)$, 
see equation~(3.5)), the other two at $H=100, 150$ 
correspond to modes $n=2,3$ (i.e., $\hat\nu\sim \cos(2\pi y)$ and $\sin(3\pi y)$, respectively).
In fact, there is an infinite hierarchy of such modes as $H\to\infty$
which has been discussed before (Alam \& Nott 1998; Alam \etal~2005).

For the dilute limit of model-$A$ (i.e., model-$A_0$),
there are {\it stationary} instabilities at finite densities ($\nu>0$),
and the stability diagram (not shown for brevity) in the ($H,\nu$)-plane looks similar to 
that for model-$A$ (figure~1), but the range of $H$ over which
the flow is unstable is much larger.
As expected, both model-A and model-$A_0$
predict that the flow is stable in the Boltzmann limit ($\nu\to 0$).

\begin{figure}
\includegraphics[width=7.0cm]{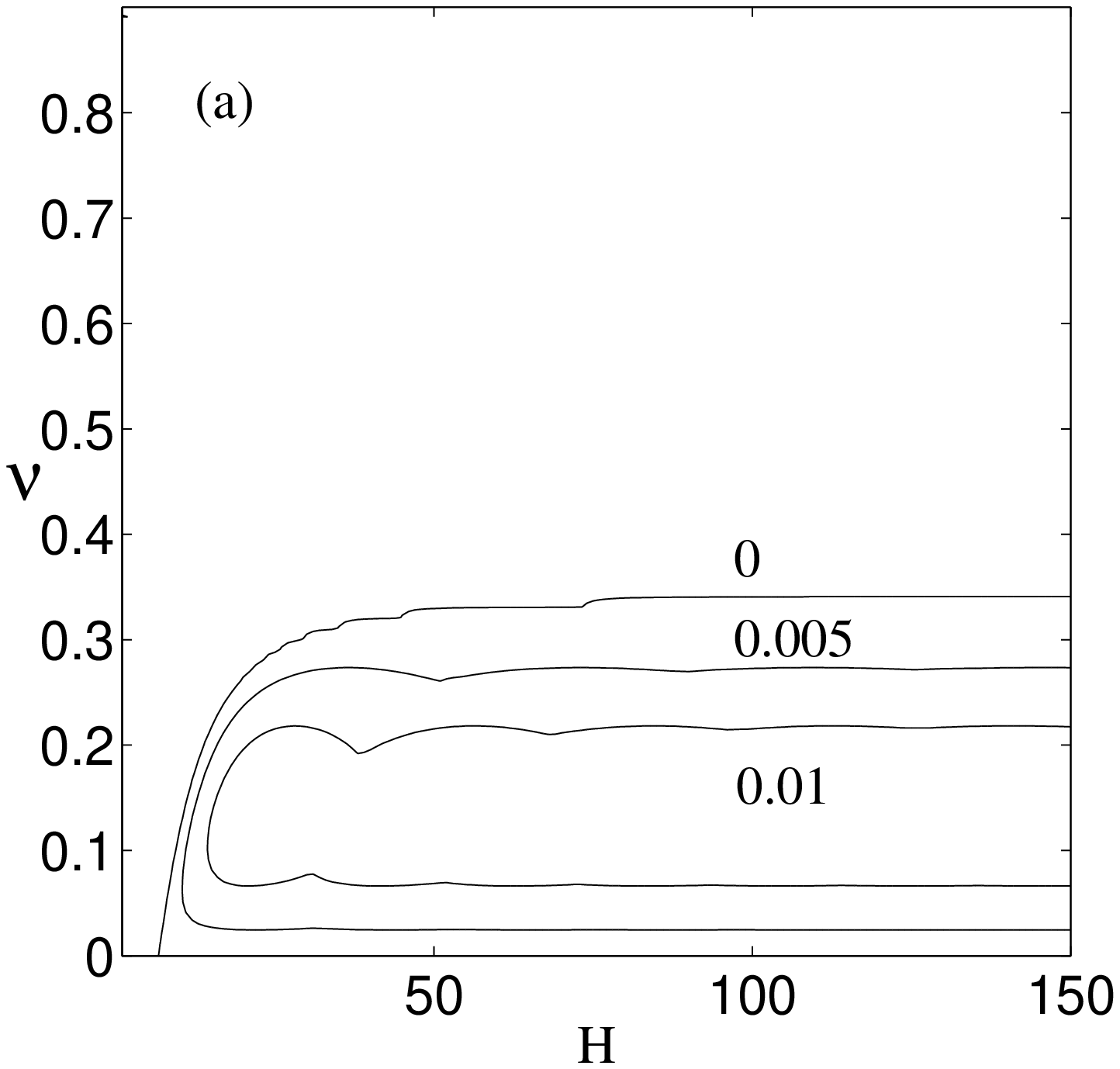}
\includegraphics[width=7.0cm]{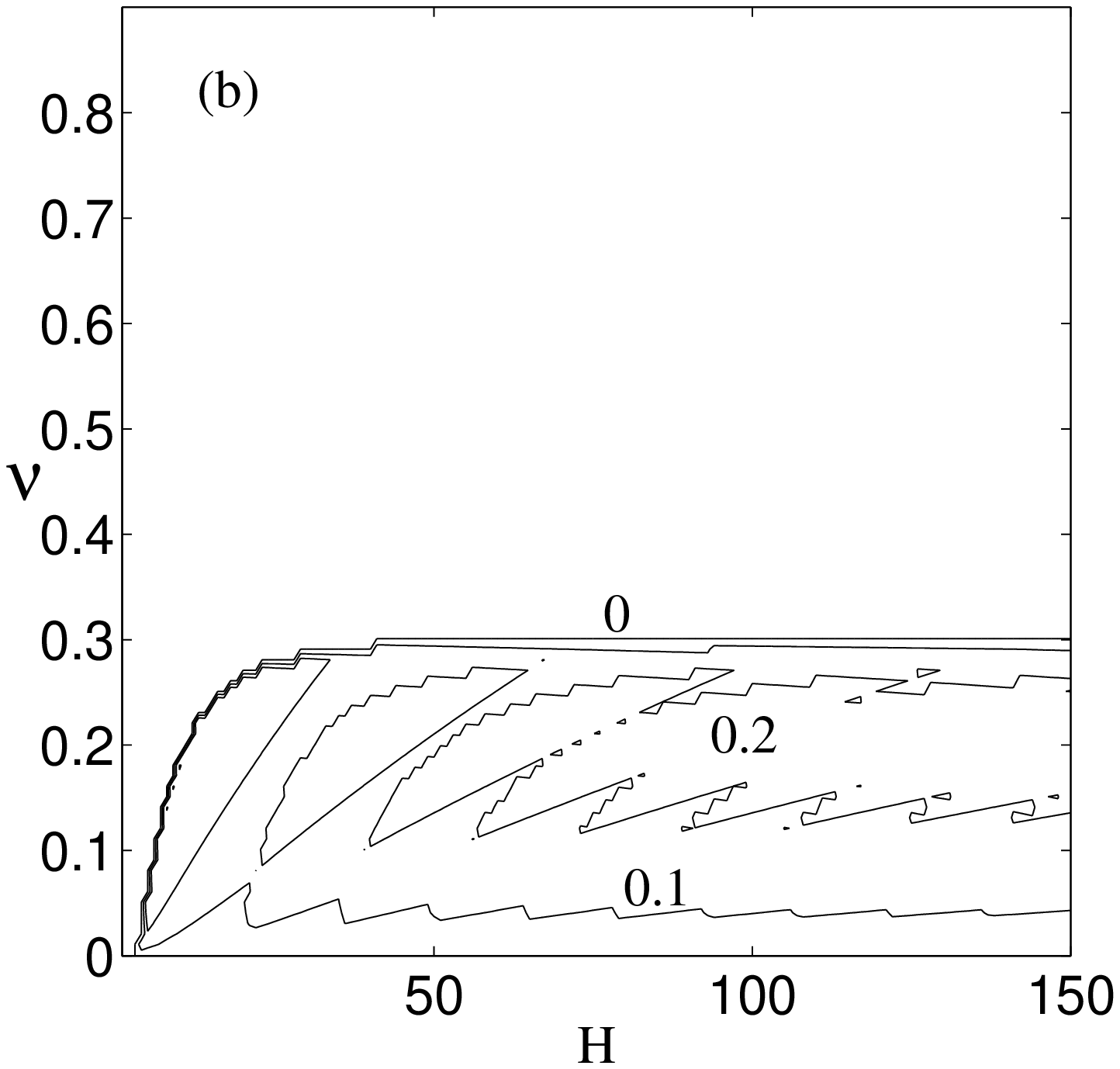}
\caption{
($a$) Phase diagram, showing the positive growth-rate contours, for  model-$A_d$ 
(i.e., the dense limit of model-A): $e=0.9$, $\nu_{m}=\pi/2\sqrt{3}$.
($b$) Contours of frequency, indicating that the instability in panel $a$ is due
to travelling waves.
}
\label{fig:fig3}
\end{figure}

\begin{figure}
\includegraphics[width=7.2cm]{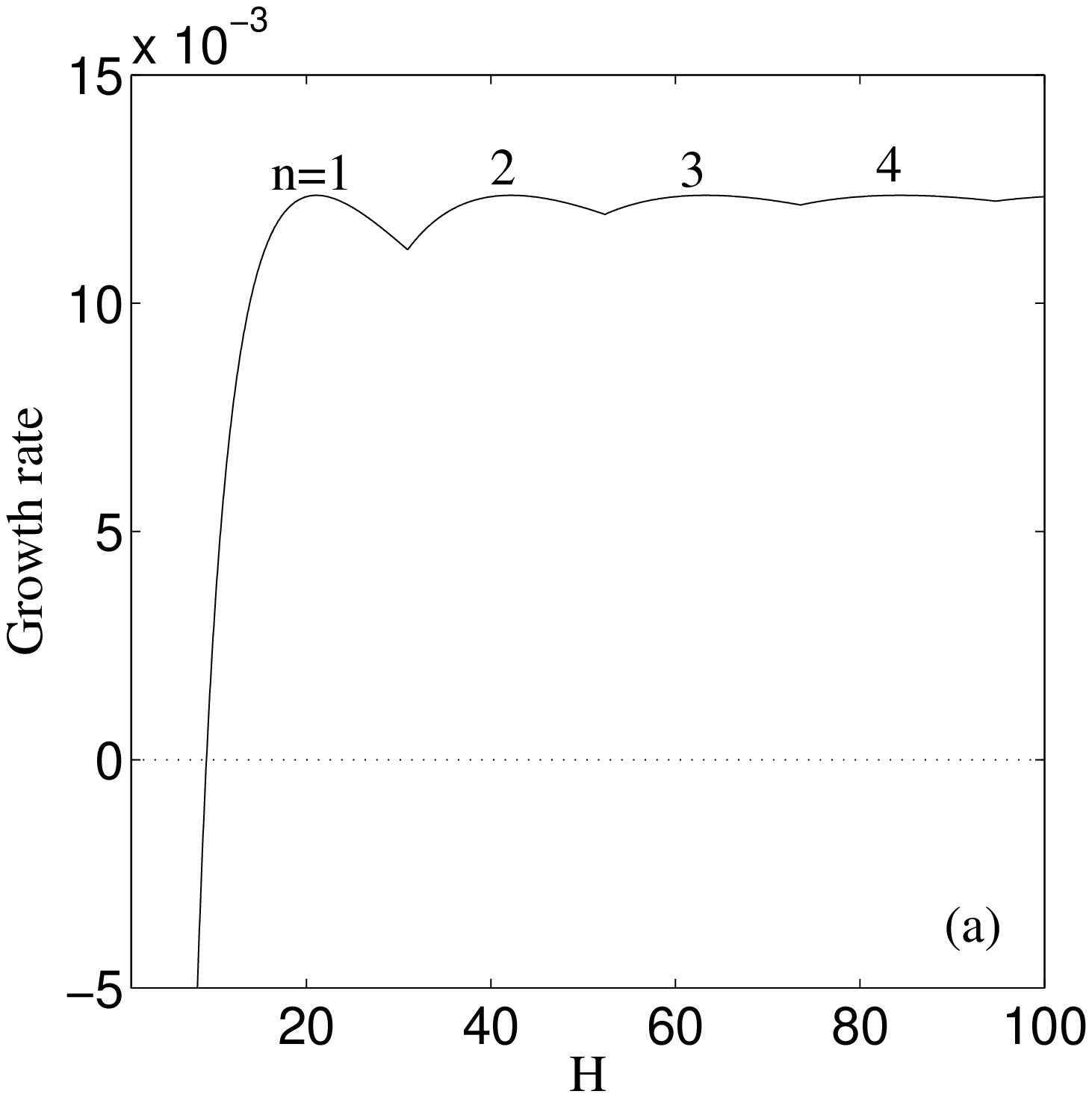}
\includegraphics[width=7.2cm]{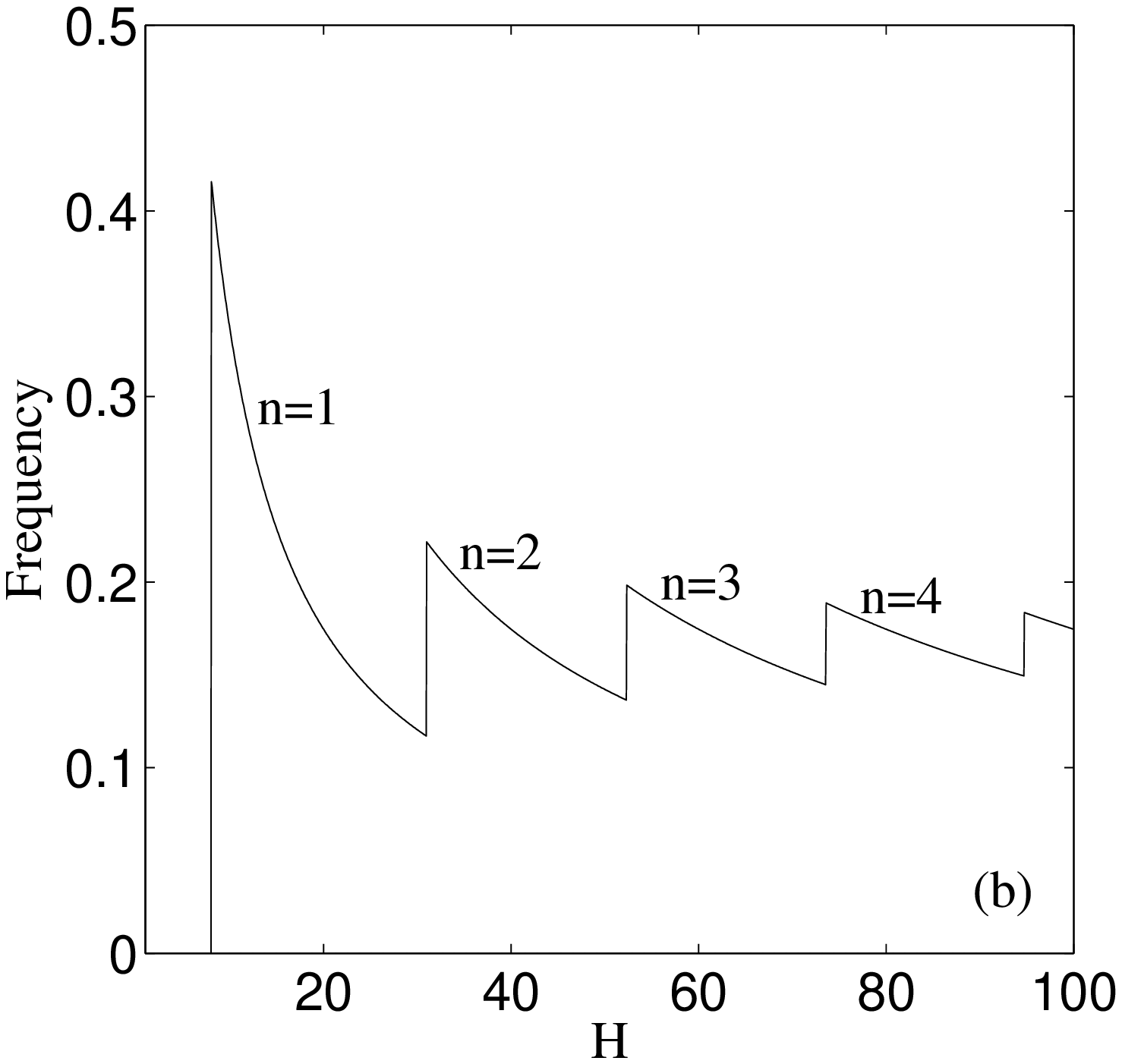}
\caption{
Variations of ($a$) the growth rate and ($b$) the frequency of the least stable mode
with $H$ for $\nu=0.1$ and $e=0.9$. Other parameters as in figure~\ref{fig:fig3}.
}
\label{fig:fig4}
\end{figure}

Figure~3($a$) shows the analogue of figure~1 for model-$A_d$ for which the
constitutive relations are expected to be valid in the dense limit
(since the constitutive relations contain the collisional part only, \S2.1.2);
figure~3($b$) shows the contours of frequency (corresponding to the
least-stable mode) in the ($H, \nu$)-plane.
The flow is unstable to {\it travelling-wave} instabilities inside the neutral contour.
For this dense model, the crossings of different instability modes
($n=1,2,\cdots$) with increasing Couette gap $H$ and their frequency variations 
can be ascertained from figure~\ref{fig:fig4}.
From a comparison between figures~1 and 3, the following differences are noted:\\
(1) Model-$A_d$ predicts that the flow is stable in the dense limit which
is in contrast to the prediction of the full model (i.e., model-$A$) 
for which the dense flow is unstable. This is a surprising  result
since the kinetic contribution to transport coefficients is small
in the dense limit, and hence both model-$A$ and model-$A_d$ are 
expected to yield similar results.  \\
(2) There is a travelling-wave (TW) instability at low densities ($\nu<0.3$)  for model-$A_d$
which is absent in model-$A$.
Since model-$A_d$ is devoid of the kinetic modes of momentum transfer and hence not
applicable at low densities, we call the TW-modes in figure~3($a$)
as ``anomalous'' modes and they vanish when both kinetic and collisional
effects are incorporated as in the full model-$A$.

One conclusion that can be drawn from the results of three variants of model-$A$
is that the choice of the constitutive model is crucial for the prediction of
shear-banding instability. We shall come back to discuss this point in \S~5.

\subsection{Results for model-$B$: influence of global equation of state}

\begin{figure}
\includegraphics[width=8.0cm]{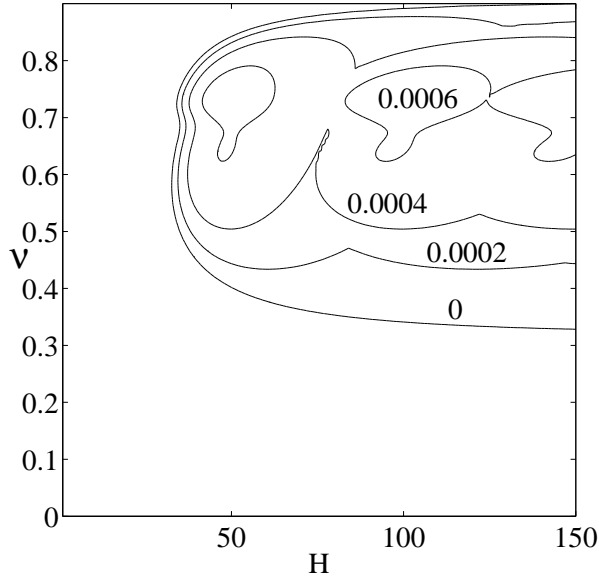}
\caption{
Phase diagram, showing the positive growth-rate contours,
for  model-$B^{\kappa}$ with a global equation for thermal conductivity: 
$e=0.9$, $\nu_{m}=\pi/2\sqrt{3}$, $\nu_f=0.7$.
}
\label{fig:fig5}
\end{figure}

Figure~5  shows a phase-diagram in the $(H, \nu)$-plane
for model $B^{\kappa}$; the flow is unstable inside the
neutral contour and stable outside.
Recall that this model is the same as model-$A$ for $\nu<\nu_f$, with the only difference being 
that we use a global equation for thermal conductivity which is valid
up-to the maximum packing density (equation 2.21).
This instability is stationary and the other features of stability diagrams remain the same 
(as those in figure~1 for the standard model-$A$), 
but there is  a dip on the neutral contour at the freezing density $\nu=\nu_f$.
For its dense counterpart, the model-$B^{\kappa}_d$  does not predict any instability
at large densities but has travelling-wave instability at low densities
(the corresponding stability diagram looks similar to that in figure~3($a$) for model-$A_d$
and hence is not shown).

\begin{figure}
\includegraphics[width=7.0cm]{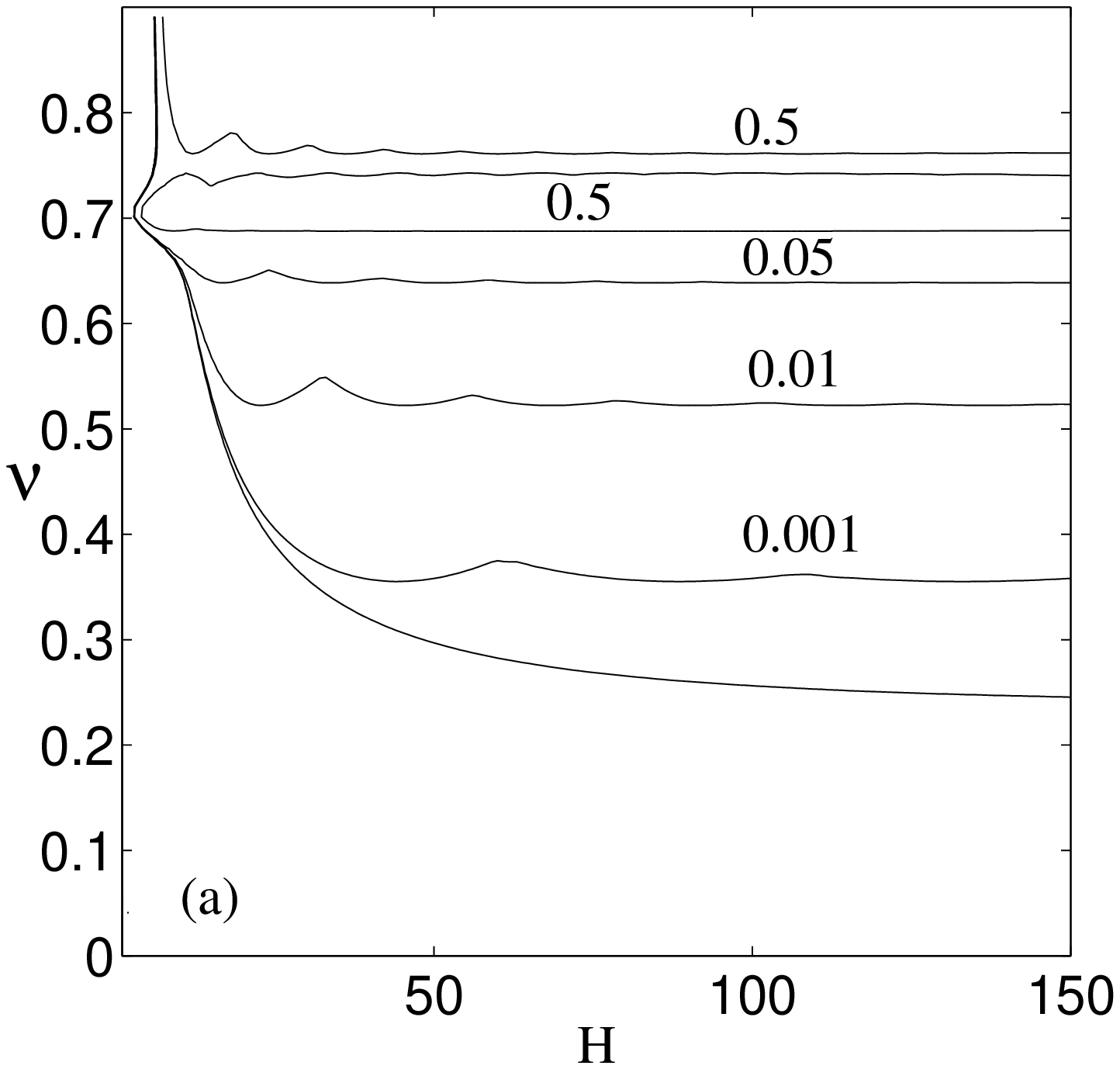}
\includegraphics[width=7.0cm]{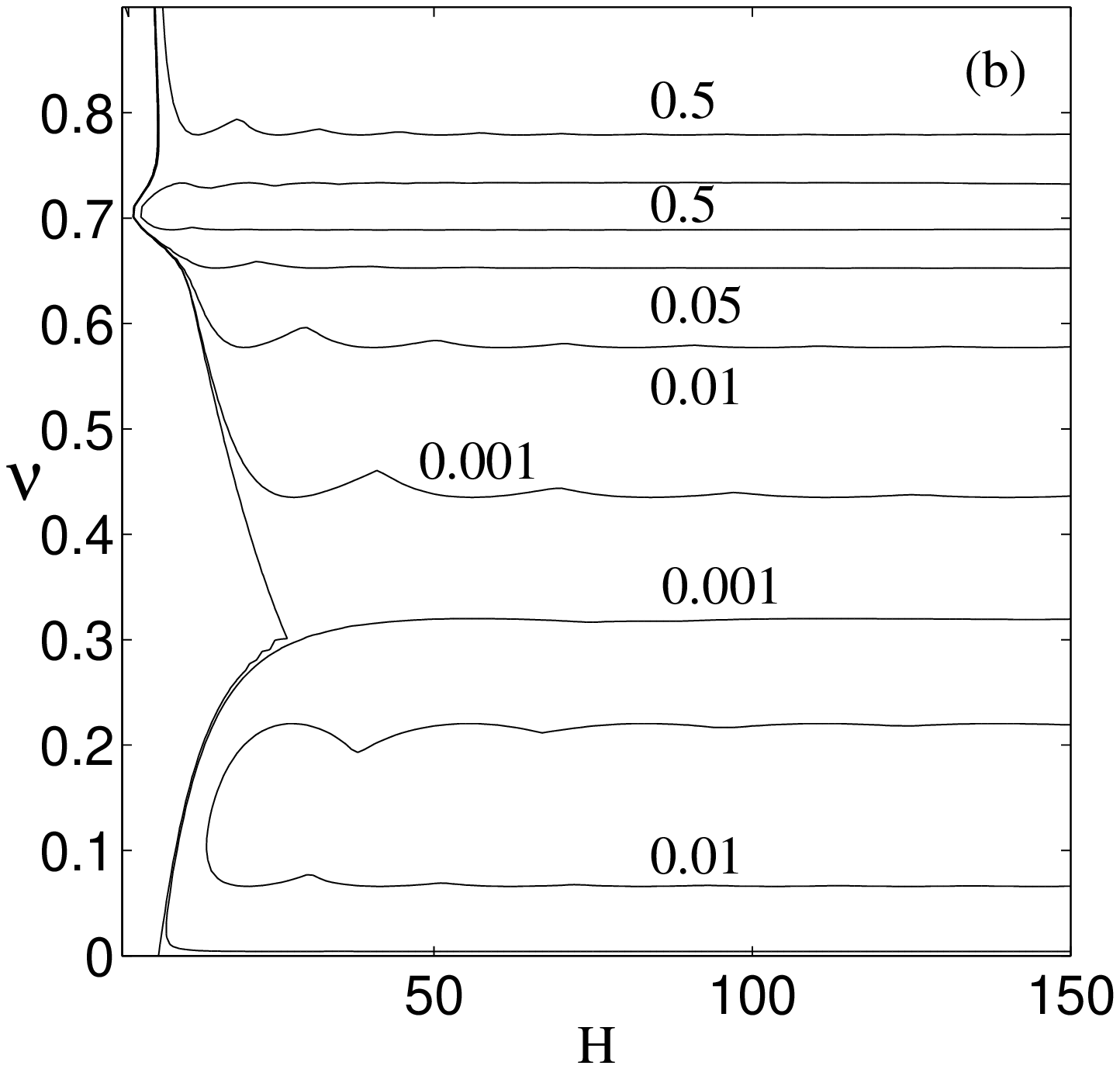}
\caption{
Phase diagram, showing the positive growth-rate contours,
for  model-$B^{p}$ with a global equation for pressure: 
$e=0.9$, $\nu_{m}=\pi/2\sqrt{3}$, $\nu_f=0.7$.
($a$) Full model;  ($b$) dense limit.
}
\label{fig:fig6}
\end{figure}

When a global equation of state for pressure is incorporated (i.e., model-$B^p$, see \S2.2),
the phase-diagram in the ($H, \nu$)-plane looks markedly different, especially at $\nu>\nu_f$,
as seen in figures~6($a$) and 6($b$).
(Recall that the model-$B^p$ is same as model-$A$, except 
that we use a global equation of state for pressure, equation (2.17).)
In figure~6($b$), the neutral stability contour contains a kink at $\nu\approx 0.3$
and there are two instability-lobes: the lower instability lobe is due to {\it travelling-waves}
and the upper-one is due to {\it stationary-waves}.
(Similar to model-$A_d$, the low density $TW$ instability in figure~6($b$) is dubbed ``anomalous''
since model-$B_d$ is not valid at low densities.)
It is interesting to note in figure~6($b$) that for the dense limit of model-$B^p$ 
(i.e., model-$B^p_d$)
the flow remains unstable to the ``stationary'' shear-banding instability
up-to the maximum packing density. This observation is
in contrast to the predictions of model-$A_d$ (figure~3$a$) and model-$B^\kappa_d$.
Therefore, we conclude that {\it within the framework of a dense model the global
equation of state for pressure induces shear-banding instabilities at large densities}.

When both the global equations for pressure and thermal conductivity are  incorporated, 
the phase diagrams in the $(H,\nu)$-plane look qualitatively similar (not shown)
to those for model-$B^p$ as in figure~6, with the only difference being
slightly higher growth rates for the least-stable mode.  
It is noteworthy that with the global EOS, the flow becomes unstable to shear-banding instabilities
at very small values of  $H$ for $\nu\ge \nu_f$.

From this section, we can conclude that within the framework of a ``dense'' 
constitutive model (that incorporates  only collisional contributions to
transport coefficients, \S2.1.2), a simple modification with a global
equation of state for {\it pressure} induces new shear-banding instabilities at large densities;
however, a similar modification with a global
equation of state for {\it thermal conductivity} does not
induce any new instability.

\subsection{Results for model-$C$ and model-$D$: influence of viscosity divergence}

\begin{figure}
\includegraphics[width=7.0cm]{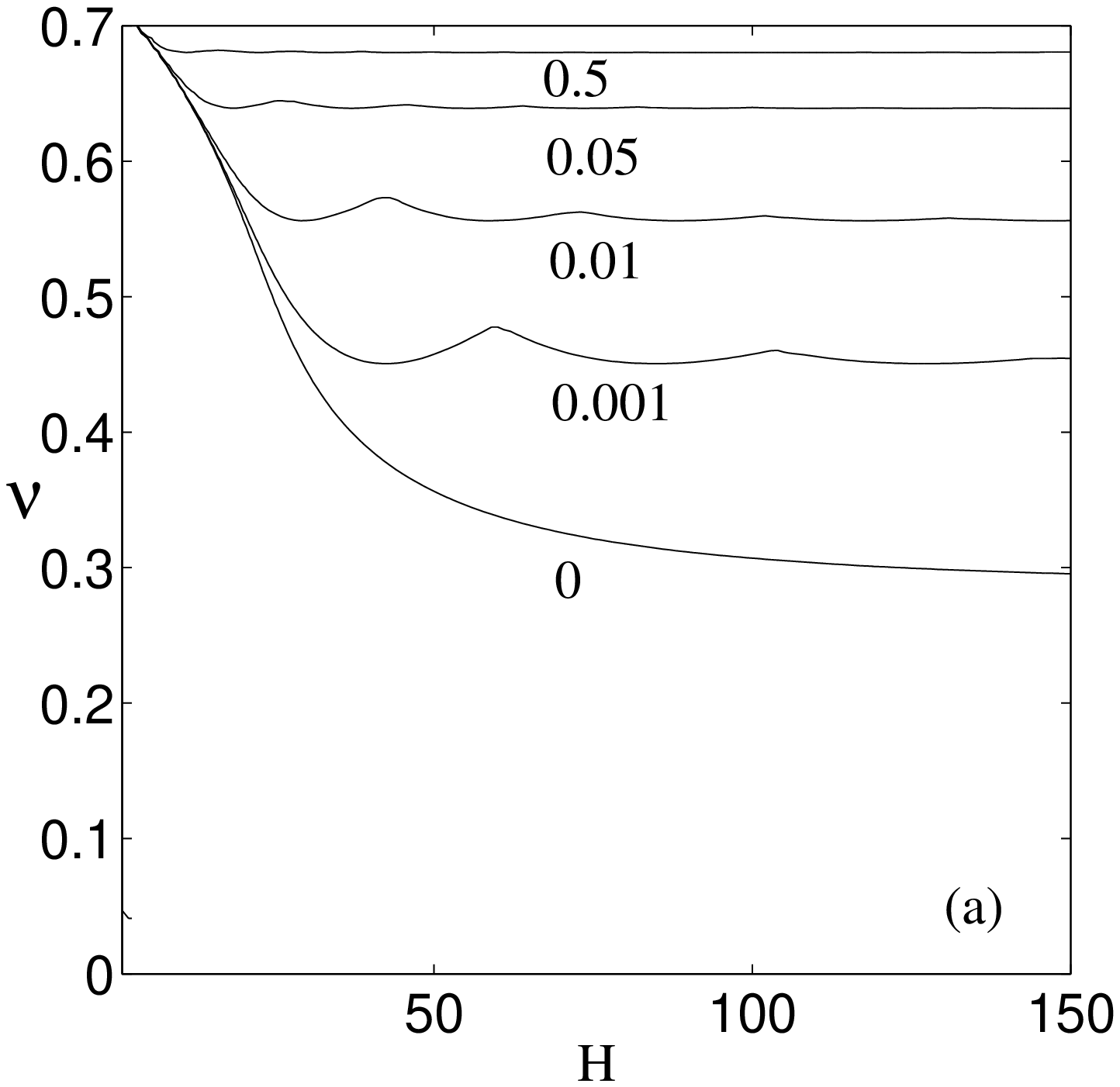}
\includegraphics[width=7.0cm]{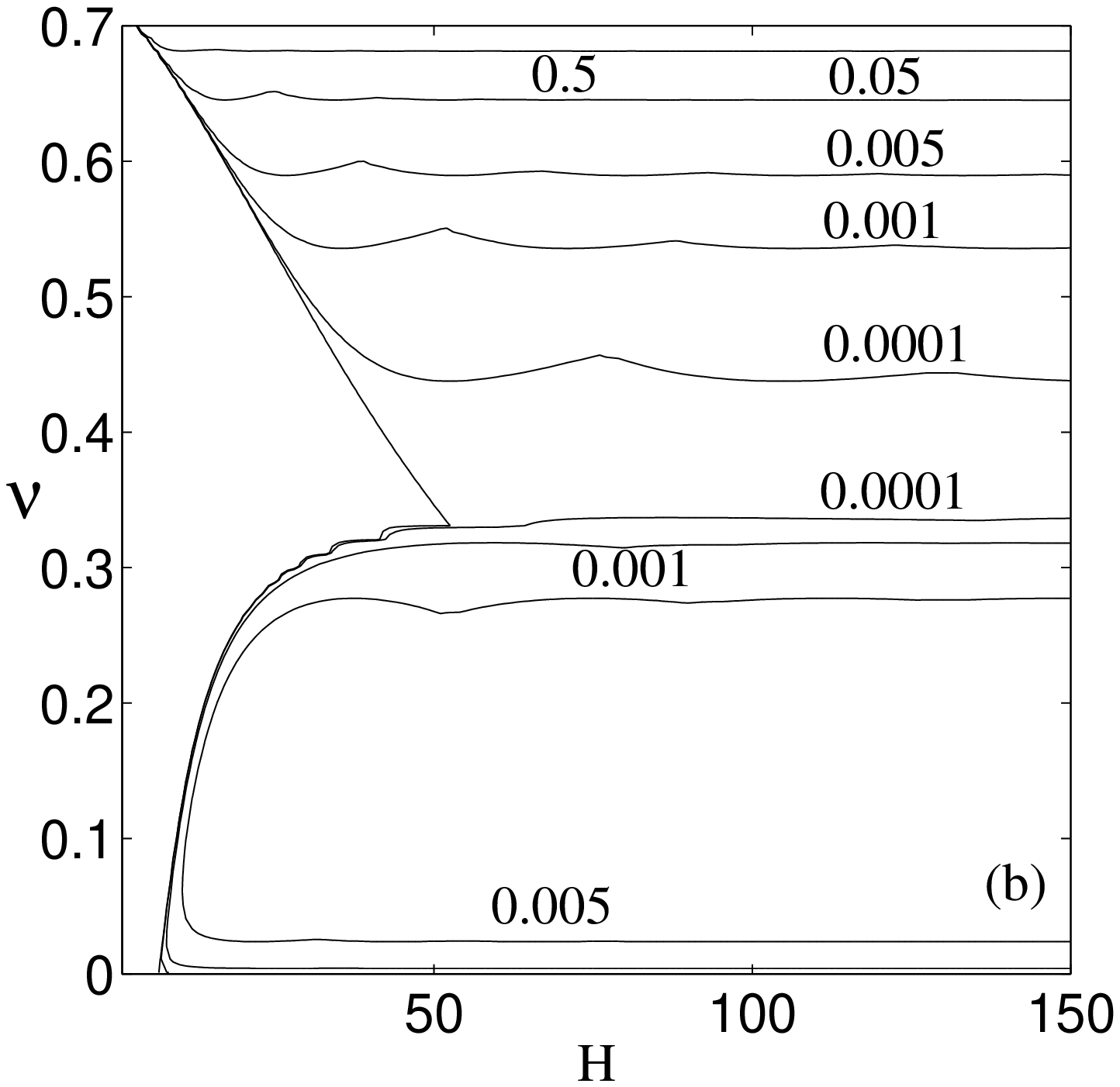}
\caption{
Phase diagram for  model-$C$ with viscosity divergence: $e=0.9$, $\nu_{m}=\pi/2\sqrt{3}$, 
$\nu_\mu=0.71$. ($a$) Full model;  ($b$) dense limit.
}
\label{fig:fig7}
\end{figure}

As discussed in \S2.3, model-$C$ is the same as model-$A$, with the
viscosity divergence being at a lower density $\nu=\nu_\mu<\nu_m$.
On the other hand,  model-D  is the most general model that incorporates 
the viscosity divergence at a  lower density $\nu=\nu_\mu<\nu_m$ (equation 2.23)
as well as global equations for pressure (equation 2.17) and thermal conductivity (equation 2.21).
The stability results for these two models are found 
to be similar, and hence we present results only for model-$C$.

Figures~7($a$) and 7($b$) show phase-diagrams in the ($H, \nu$)-plane for model-$C$ and 
its dense variant model-$C_d$,
respectively; the results are shown up-to the viscosity divergence since uniform shear is not
a solution for $\nu>\nu_\mu$.
For the full model in figure~7($a$), the phase-diagram looks similar to those 
for model-$A$ and model-$B$.
For its dense counterpart in figure~7($b$), the phase-diagram is similar to that for model-$B_d$
and model-$B^p_d$ in the sense that all three models support shear-banding instabilities at large
densities.   Therefore,  for model-$C_d$, the viscosity divergence  induces shear-banding
instabilities at large densities.  
This prediction is in tune with the  results of Khain \& Meerson (2006)
whose model is similar to our model-$C_d$ (see \S~5.3).

Within the framework of a ``dense'' model (\S2.1.2), therefore,  we can conclude
about the emergence of shear-banding instabilities at large densities:
(1) a global equation of state for pressure alone can induce shear-banding instabilities;
(2) a  viscosity-divergence at a density, $\nu=\nu_\mu$, lower
than the maximum packing density  alone can induce shear-banding instabilities.

\section{Discussion: Universality of shear-banding instability and crystallization}

\subsection{An universal criterion for shear-banding instability}

Since the shear-banding instability is a {\it stationary} ($\omega_i=0$) mode,
this instability is given by  one of the real roots (equation~(3.10)) of the dispersion relation.
The condition for neutral stability ($\omega_r=0$) can be  obtained by setting $\omega=0$ in
the dispersion relation (\ref{eqn_dispersion1}):
\begin{equation}
    a_0=0 \quad \Rightarrow \quad  k_n^2/H^2  =  \frac{\Psi_2}{\Psi_1},
\label{eqn_neutral1}
\end{equation}
where
\[
   \Psi_1 = \frac{f_4}{f_5}
   \quad
   \mbox{and}
   \quad
   \Psi_2 = \left(\frac{f_{5\nu}}{f_5} + \frac{f_{2\nu}}{f_2}\right)\frac{f_1}{f_{1\nu}} -2 .
\]
For an instability to occur at a given density, there 
must be a  range of `positive' discrete wave-numbers, i.e., $k_n/H>0$,
which is equivalent to
\[
  \Psi_2 >0,
\]
since $\Psi_1$ is always positive.  
The expression for $\Psi_2$ can be rearranged to yield (Alam 2006)
\begin{equation}
   \frac{\rm d}{{\rm d}\nu}\left(\frac{\sqrt{f_2 f_5}}{f_1}\right) >0,
\quad\quad (\mbox{with}\;\; f_{1\nu}>0)
\label{eqn_criterion}
\end{equation}
which must be satisfied for the onset of instability.  
(It should be noted that (\ref{eqn_criterion}) provides a necessary
condition for instability, but the sufficient condition is
tied to the thermal-diffusive mechanism that leads to
an instability length scale (\ref{eqn_neutral1}) which is discussed in \S5.2.)
The term within the bracket in (\ref{eqn_criterion})
is the ratio between the shear stress, $P_{xy}=\mu\gamma$, 
and the pressure, $p$, for the plane Couette flow:
\begin{equation}
   \beta_d = \frac{P_{xy}}{p} = \frac{\mu(\frac{{\rm d}u}{{\rm d}y})}{p} 
     = \frac{f_2\sqrt{T}}{f_1 T} = \frac{f_2}{f_1 \sqrt{T}}
     = \frac{f_2}{f_1 \sqrt{f_2/f_5}} \equiv \frac{\sqrt{f_2 f_5}}{f_1},
\label{eqn_criterion1}
\end{equation}
where $\gamma={\rm d}u/{\rm d}y$ is the local shear rate
(which is a constant for uniform shear flow).
This is nothing but the {\it dynamic friction coefficient} of the shear flow,
which must increase with increasing density for the shear-banding instability to occur.
Note that as per Navier-Stokes-level description, the steady fully developed plane
Couette flow admits solutions for which the shear stress and pressure are constants 
across the Couette gap.
Hence, the  dynamic friction coefficient, $\beta_d=\mu({\rm d}u/{\rm d}y)/p = \mu\gamma/p$, 
is a {\it position-independent}
order-parameter for both ``uniform'' ($\gamma=const.$) and ``non-uniform'' ($\gamma=\gamma(y)$)
shear flows.

\begin{figure}
\includegraphics[width=8.0cm]{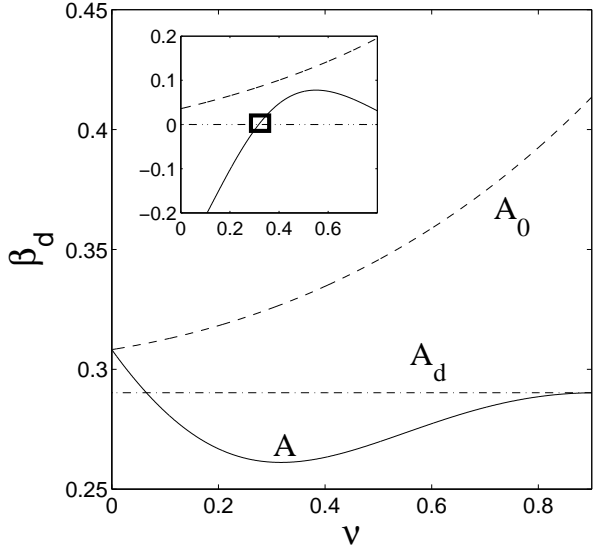}
\caption{
Variation of the ``dynamic'' friction coefficient, $\beta_d=\mu\gamma/p$, with density
for model-$A$: $e=0.9$, $\nu_{m}=\pi/2\sqrt{3}$.
The inset shows the variation of $\frac{d\beta_d}{d\nu}$ with $\nu$:
the onset of shear-banding instability corresponds to $\frac{d\beta_d}{d\nu}>0$, denoted
by the square-symbol.
}
\label{fig:fig8}
\end{figure}

For model-A, the variation
of $\beta_d$ with $\nu$ is non-monotonic as shown by the solid 
line in figure~8:  in  the dilute limit $\beta_d$ is large whose value
decreases with increasing density till a critical density $\nu=\nu_{cr}$
is reached beyond which $\beta_d$ increases. 
Recall that the onset of shear-banding instability corresponds to $\frac{d\beta_d}{d\nu}>0$, 
denoted by the square-symbol in the inset of figure~8.
It is clear from this inset that the full model is unstable to 
shear-banding instability for  $\nu>\nu_{cr}$,
the dilute model is unstable at any density  (since $\frac{d\beta_d}{d\nu}>0$ for $\nu>0$),
but the dense model is stable for all densities (since $\frac{d\beta_d}{d\nu}=0$).
It is worth pointing out that, for the full model-$A$,
the critical density for the onset of the shear-banding instability (cf. figure~1) is
$\nu_{cr}\approx 0.31733$ which is independent of the restitution coefficient $e$
as confirmed in figure~9. 
The lower branch of the neutral contour of figure~1 asymptotically approaches this critical density
as $H\to\infty$.

\begin{figure}
\includegraphics[width=8.0cm]{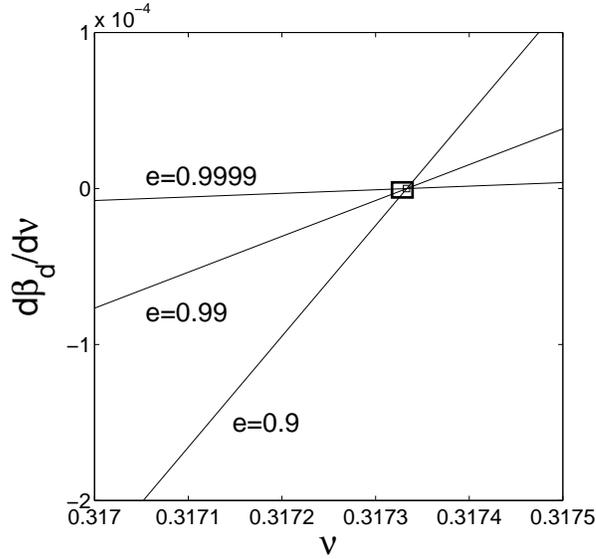}
\caption{
The effect of restitution coefficient $e$ on  the variation of $\frac{d\beta_d}{d\nu}$ with $\nu$
for model $A$.
}
\label{fig:fig9}
\end{figure}

\begin{figure}
\includegraphics[width=8.0cm]{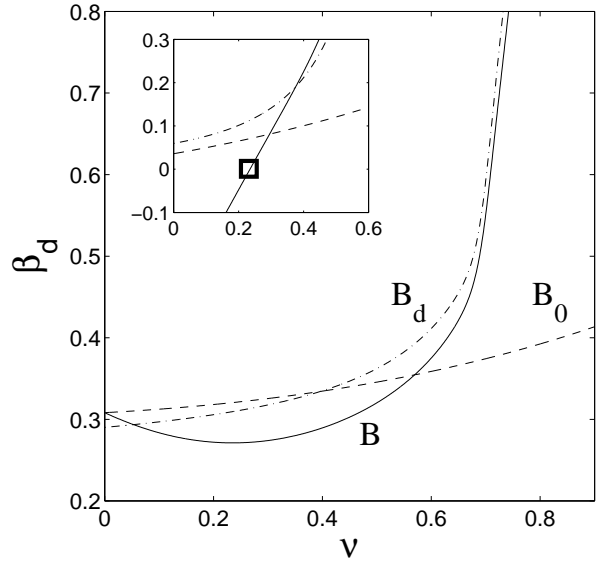}
\caption{
Same as figure~\ref{fig:fig8}, but for model $B$.
}
\label{fig:fig10}
\end{figure}

For model-$B$, the variation of $\beta_d$ with density ($\nu$) is shown in figure~10,
and the inset shows the variation of $d\beta_d/d\nu$ with $\nu$.
(Recall that this model
incorporates the global equations of states for pressure and thermal conductivity.)
In contrast to the `stable' model-$A_d$, the model-$B_d$ (i.e. the dense variant of model-$B$)
is unstable to shear-banding instabilities for all densities (figure~6$b$)
since $\frac{d\beta_d}{d\nu}>0$.  For the full model-$B$,
the critical density for the onset of shear-banding instability (cf. figure~6$a$) is
$\nu_{cr}\approx 0.2351$ which is also independent of the restitution coefficient $e$
(as in figure~9 for model-$A$).
For models $C$ and $D$, this critical density is
$\nu_{cr}\approx 0.2849$ and $\nu_{cr}\approx 0.2207$, respectively.

It should noted that there is a constraint ($f_{1\nu}>0$) with  our  instability criterion (5.2),
i.e., $f_1$ should be a monotonically increasing function of density for
the instability to occur.
This constraint on $f_1$, see equations (2.7) and (2.19), is satisfied for all four models.
The consequence of a possible non-monotonic $f_1$ is that the shear flow remains
stable over a small density range between freezing and melting.
However, the increasing dynamic friction with increasing density (5.2) still remains
the criterion for the onset of the shear-banding instability.
Even though for an unsheared elastic system, $f_1$ is non-monotonic (Luding 2001)
between freezing and melting density, we retain equation (2.19-2.20) for $f_1$
since this functional form has been shown to hold up-to the random close-packing density
in simulations of plane shear flow of inelastic hard-disks (Volfson \etal~2003).

\subsubsection{Shear-banding state and crystallization}

For all four models ($A$, $B$, $C$ and $D$), the shear flow is stable
and the system remains homogeneous at low densities, but becomes unstable
to shear-banding instability beyond a critical density ($\nu>\nu_{cr}$). 
This is due to the fact that for $\nu>\nu_{cr}$
the flow cannot sustain the  increasing dynamic friction ($\beta_d$) and hence breaks
into alternating layers of dilute and dense regions along the gradient direction.
For $\nu>\nu_{cr}$, the associated ``finite-amplitude'' bifurcated  solution corresponds to a lower 
shear stress, or, equivalently, a lower dynamic friction coefficient.
This has been verified (Alam 2008) numerically by tracking the bifurcated  solutions
of the associated steady nonlinear equations.

A representative set of such nonlinear bifurcated solutions 
for the profiles of density $\nu(y)$, granular temperature $T(y)$
and stream-wise velocity $u(y)$ are displayed in figure~11 for three values
of the Couette gap $H=50$, $75$ and $125$ for model-$A$;
the related numerical procedure is the same as  described in  Alam \etal~(2005). 
For this plot, the mean density
is set to $0.5$ and the restitution coefficient is $0.9$; the corresponding
growth-rate variation of the least stable mode can be ascertained from figure~2($a$).
These nonlinear solutions bifurcate from  the $n=1$ mode (see, equations 3.5 and 3.6)
of the corresponding linear stability equations,
and there is a pair of nonlinear solutions for each $H$ due to the
symmetry of the plane Couette flow (Alam \& Nott 1998).
For mode $n=1$, the density is maximum at either of the two walls, and 
this density-maximum approaches the maximum packing density 
($\nu_m\approx 0.906$) at $H=125$ (solid line in figure~11$a$)
for which we have the coexistence of a ``crystalline'' zone and a dilute zone,
representing a state of phase separation.
Within the crystalline zone, the granular temperature approaches zero (figure~11$b$)
and so does the shear rate (figure~11$c$).
It is noteworthy that the shear-rate  is almost uniform and localized within the dilute zone.
The resulting two-phase solution is 
called a ``shear-band'' (or, shear-localization) since the shearing
is confined within a band of an agitated dilute region that coexists with a denser region 
with negligible shearing (i.e., a crystalline-region at large enough Couette gap).

At $H=50$ with parameters as in figure~11,
there are two possible solutions: a ``uniform-shear'' state,
with the ``dynamic'' friction coefficient being $\beta_d\approx 0.26965$,
which is unstable, and one of two ``shear-band'' solutions
for which $\beta_d\approx 0.2672$.
The selection of the stable branch is  determined by
the value of the dynamic friction coefficient being the lowest among
all possible solutions, and
therefore the equilibrium state of the flow (at $H=50$) corresponds to
the shear-banding state of a ``lower'' dynamic friction (Alam 2008).

For higher-order modes ($n=2,3,\cdots$), the shape of  
the nonlinear density/temperature/velocity profiles
can be ascertained from (3.5) and (3.6).
For example, the density profiles for modes $n=2$ and $3$
would look like the corresponding density eigenfunctions in figure~2($b$).
In fact, the solution for the first mode ($n=1$)
serves as a ``building-block'' of solutions for higher modes
which  has been clarified previously (Alam \& Nott 1998; Nott \etal~1999; Alam \etal~2005).

At this point, we can say that an accurate constitutive model
over the whole range of densities (that incorporates both kinetic 
and collisional modes of transport mechanisms) should be used
since the dilute and  dense regimes can coexist even at a moderately low mean density.
Details of such shear-banding solutions for other models ($B$, $C$ and $D$)
and their stability as well as the related results from 
particle-dynamics simulations will be considered in a separate paper.
Such an exercise will help to identify and tune the best among the four models.

\begin{figure}
\includegraphics[width=7.0cm]{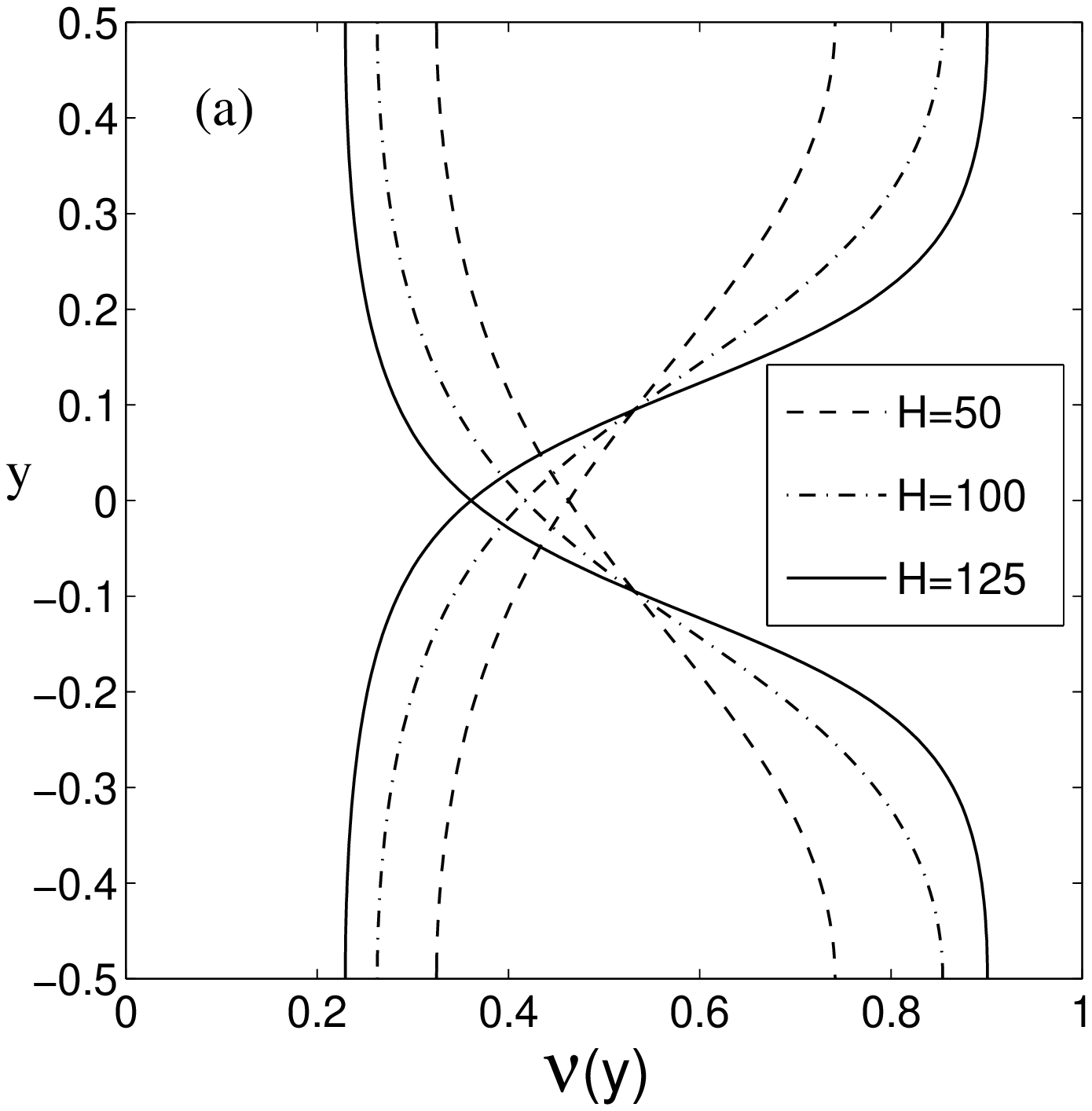}
\includegraphics[width=7.0cm]{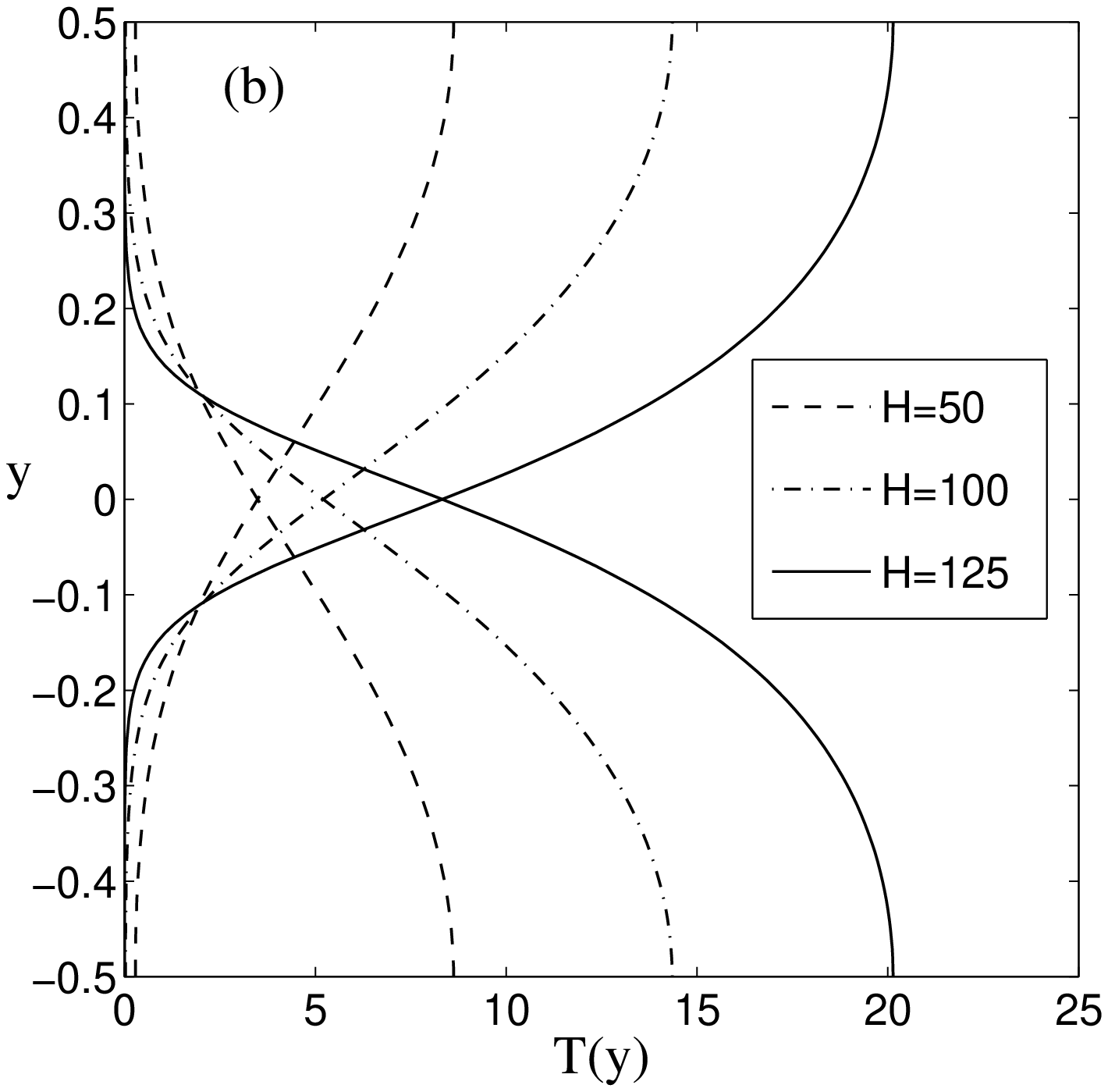}
\includegraphics[width=7.0cm]{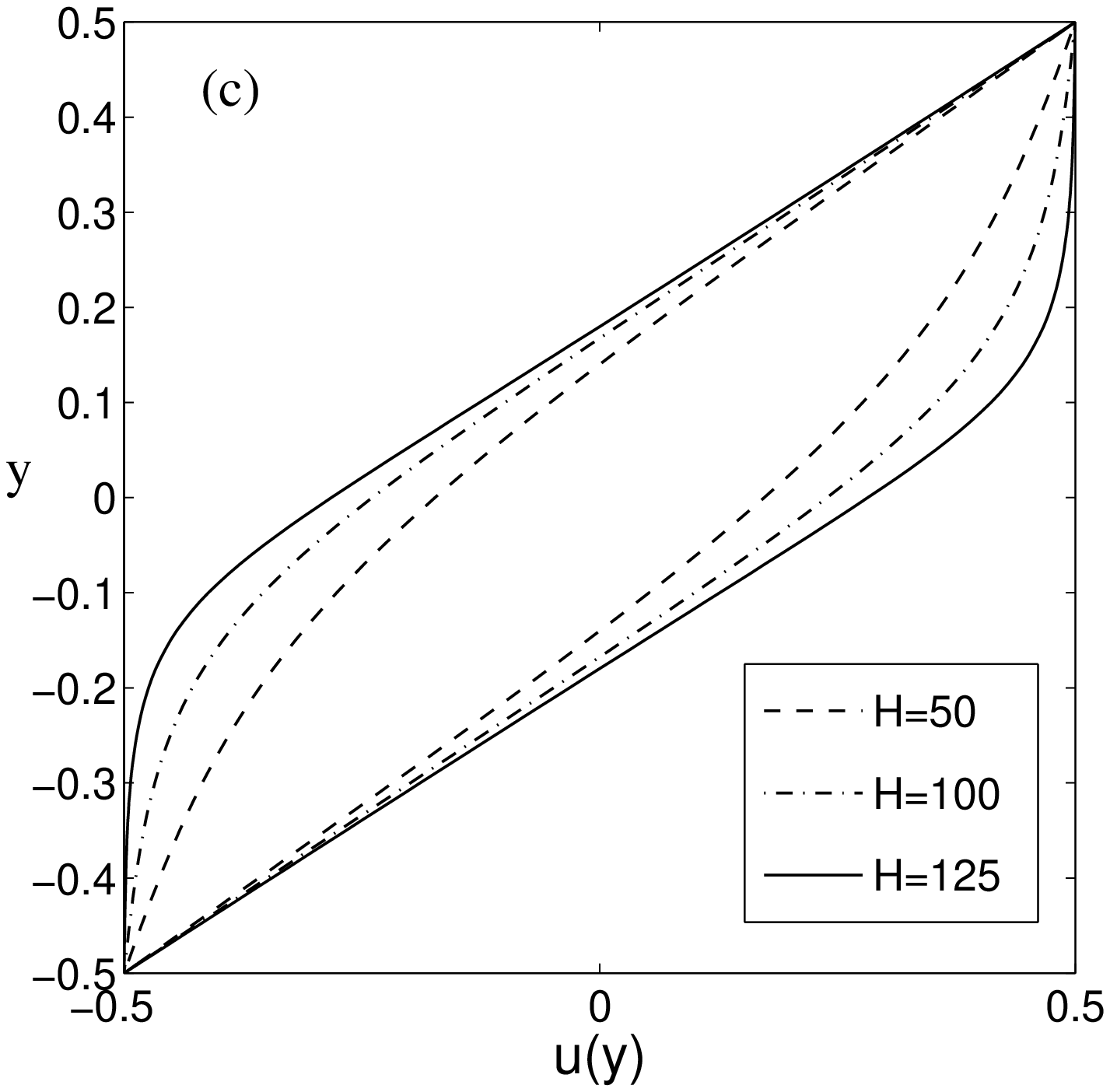}
\caption{
Nonlinear shear-banding solutions for model-$A$ for mode $n=1$ with  $\nu=0.5$ and  $e=0.9$:
($a$) density, ($b$) granular temperature and ($c$) stream-wise velocity.
}
\label{fig:fig11}
\end{figure}

\subsection{Instability length scale and the effect of dissipation}

While our instability criterion, given by equation~(\ref{eqn_criterion}), 
yields a critical value for the density above which the uniform shear flow
is unstable to the shear-banding instability, 
it does not say anything about the related instability length scale
below which the flow is stable (cf. figures~1--7).
This issue of a dominant instability length scale is tied to the underlying
diffusive mechanisms in a granular fluid, offered by the pseudo-thermal conductivity
in the energy balance equation (2.3).

We have mentioned in \S4 that the neutral contour 
in the ($H, \nu$)-plane (such as in figures~1--7) shifts towards right
with increasing restitution coefficient ($e$), 
and hence the flow becomes more stable in the elastic limit.
In fact, the dependence of the neutral contour on $e$ can be removed
if we define a normalized  Couette gap  as
\begin{equation}
   H^* = H\sqrt{1-e^2} ,
\label{eqn_ins_scale}
\end{equation}
which can be thought of as an ``instability length scale''
(Tan \& Goldhirsch 1997; Alam \& Nott 1998).
This length scale appears directly from an analysis of the equation for the
neutral contour (\ref{eqn_neutral1}):
\begin{eqnarray}
   H^2/\Psi_1 &=& k_n^2/\Psi_2 =f_5 H^2/f_4 = f_{50}(1-e^2)H^2/f_4 \nonumber \\
\Rightarrow  H\sqrt{1-e^2} & \equiv & H^* = k_n \sqrt{\frac{f_4(\nu)}{\Psi_2(\nu)f_{50}(\nu)}},
\label{eqn_ins_scale1}
\end{eqnarray}
where $f_5(\nu,e)=(1-e^2)f_{50}(\nu)$, and $f_4(\nu)$ is related to 
pseudo-thermal conductivity as in (2.6).
This  specific functional dependence of the ``instability
length scale'' on the restitution coefficient
is also due to the dependence of the thermal conductivity on the granular temperature
which implicitly depends on $e$ (equation~(3.2)): $T \sim f_2/f_5 \sim (1-e^2)^{-1}$.

\begin{figure}
\includegraphics[width=7.0cm]{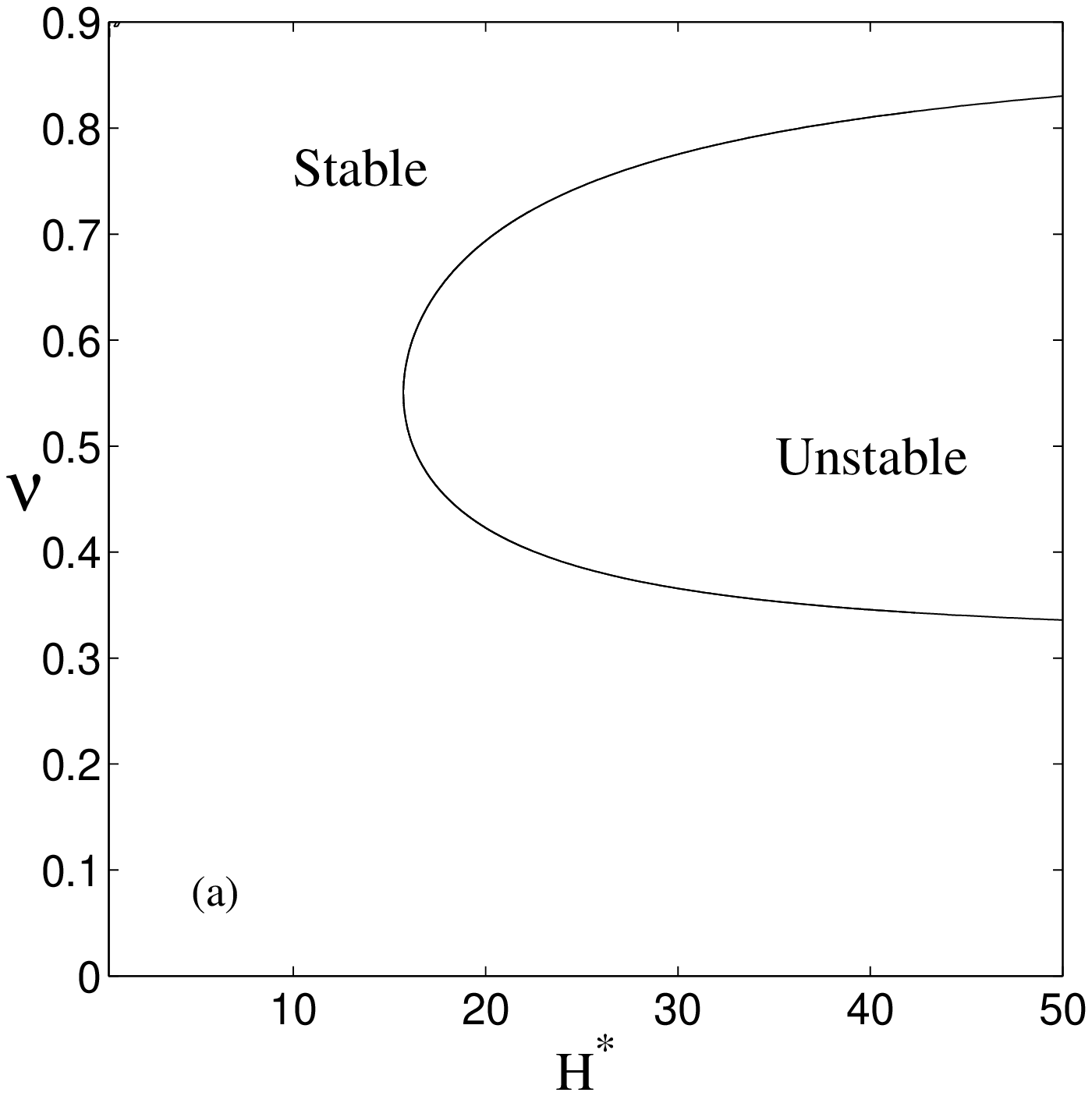}
\includegraphics[width=7.0cm]{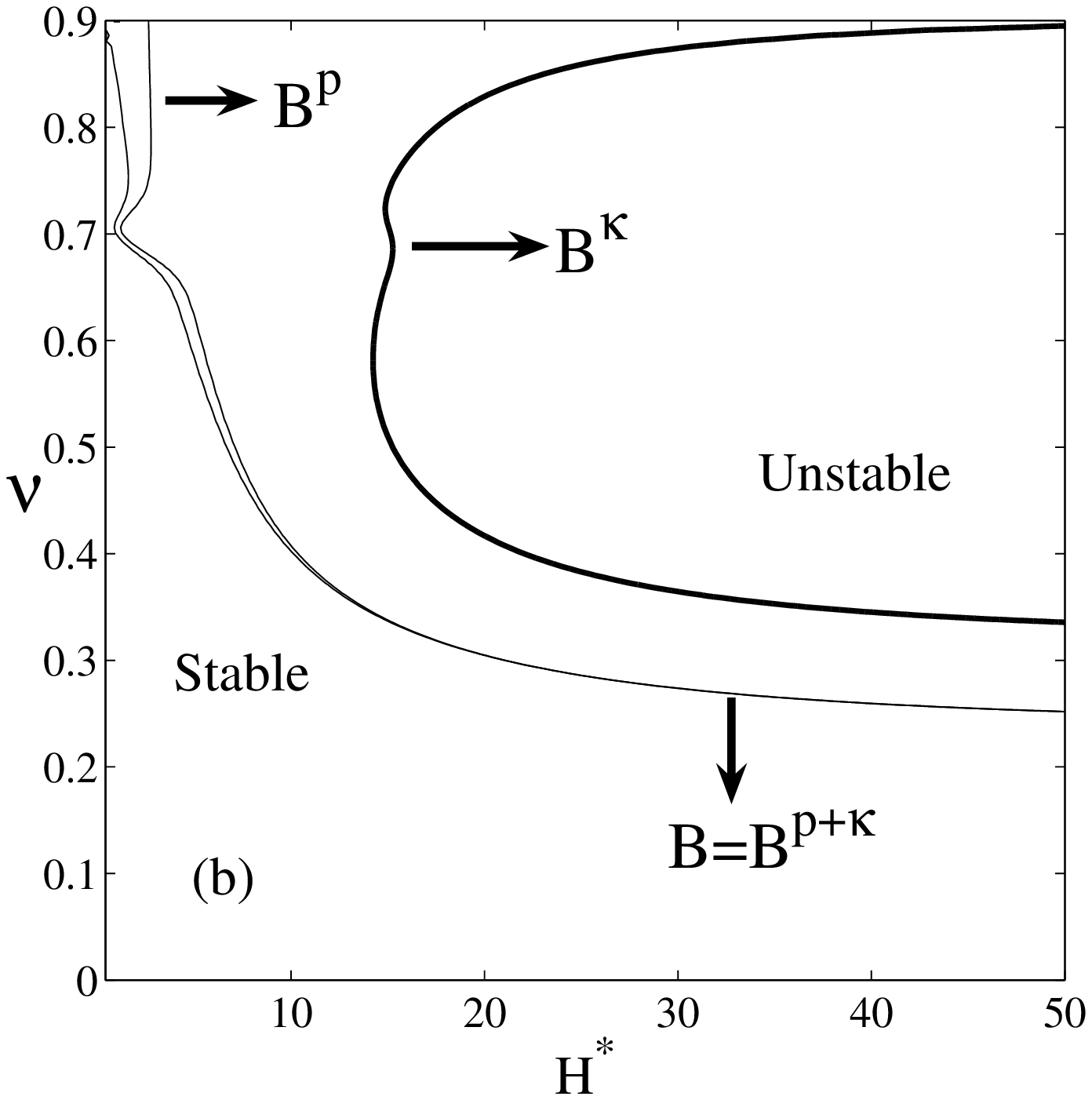}
\includegraphics[width=7.0cm]{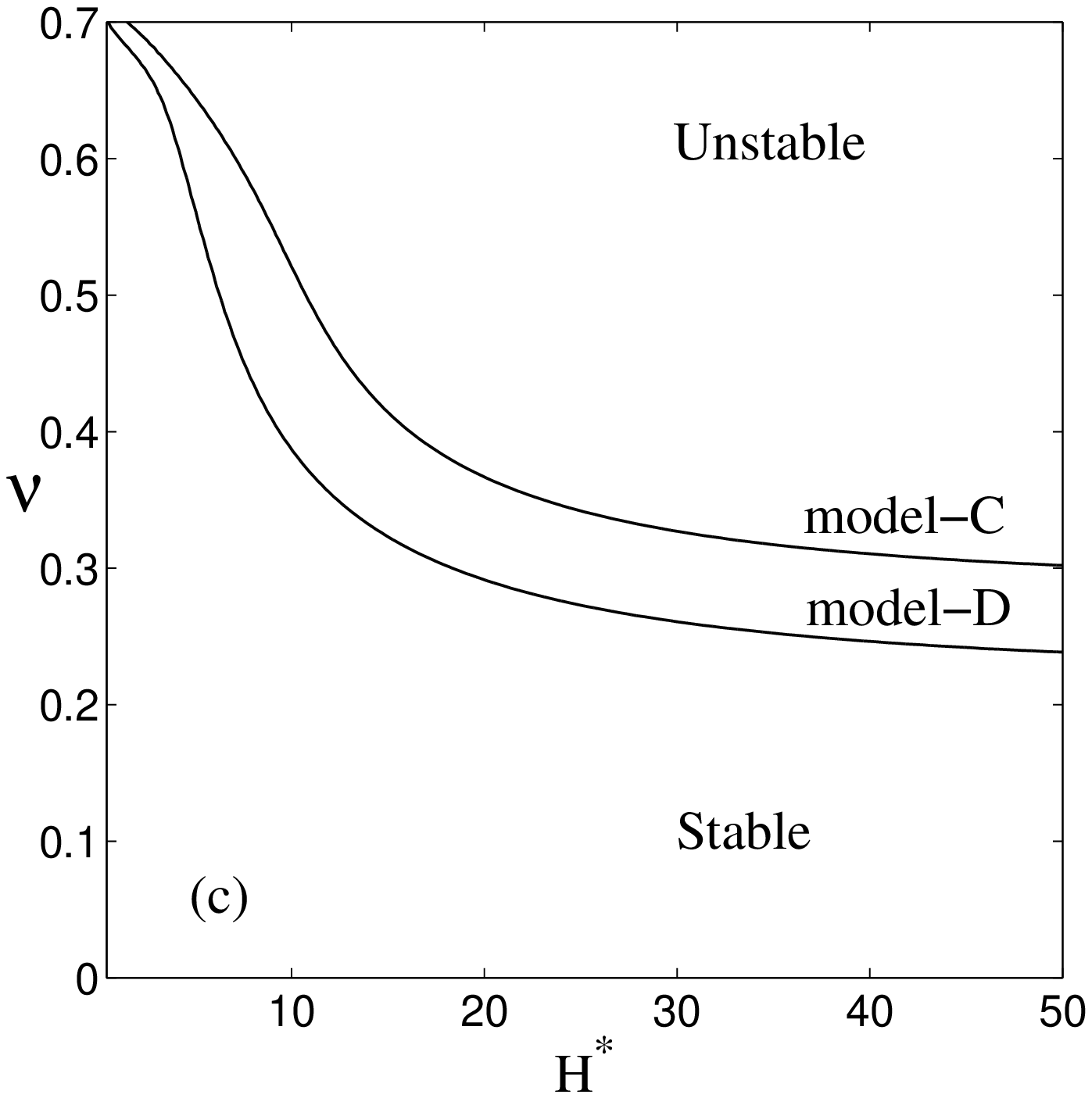}
\includegraphics[width=7.0cm]{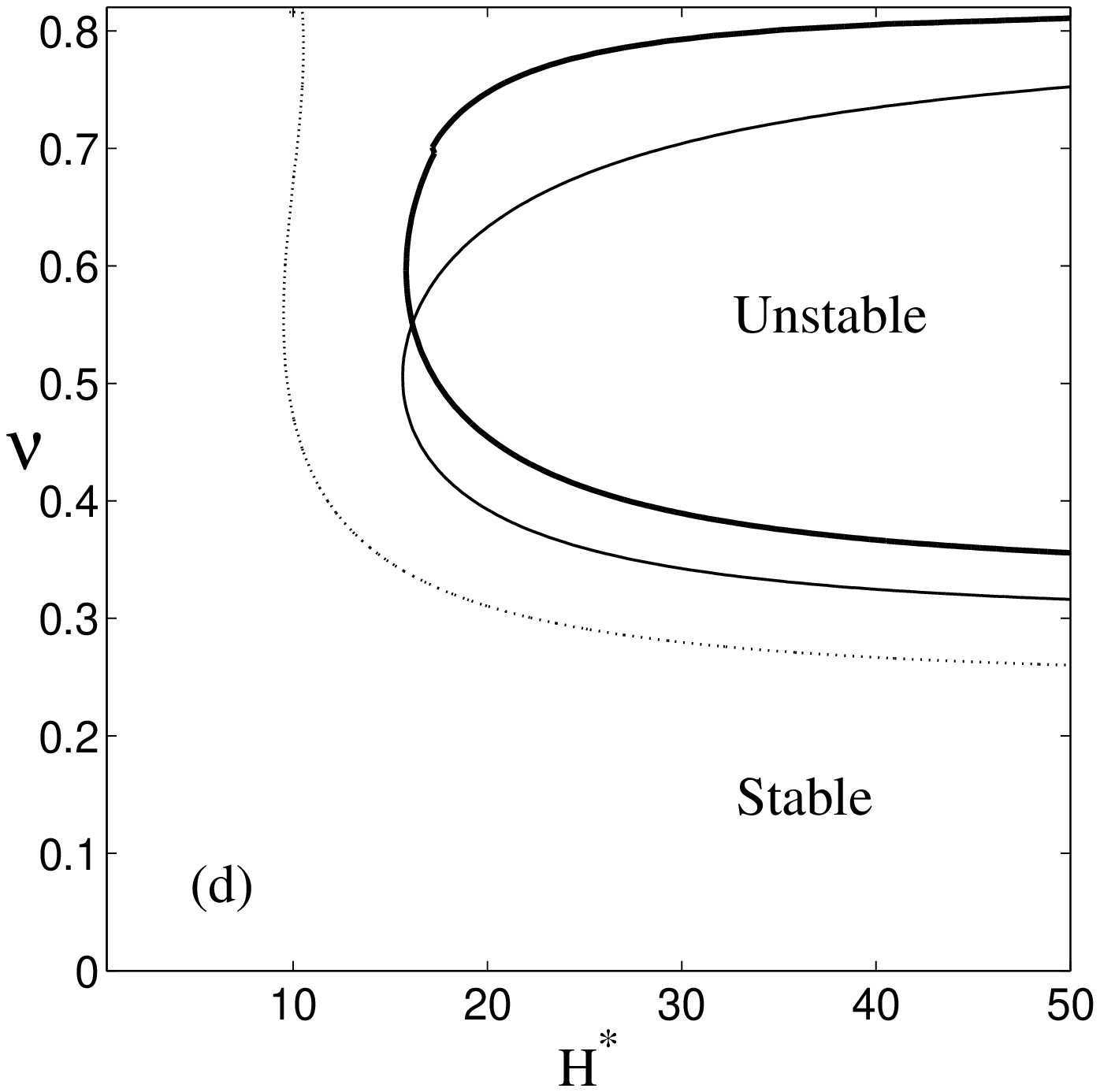}
\caption{
Renormalized stability diagrams in the ($H^*, \nu$)-plane,
showing the neutral contour that separates stable and unstable regions:
($a$) model $A$;  ($b$) model $B$;  ($c$) model $C$ and $D$;  
($d$) model $A$ with different  $\chi(\nu)$ as explained  at the end of \S5.2.
The outermost dotted curve in panel ($d$) is discussed at the end of \S5.3. 
In each panel, the abscissa has been renormalized as $H^*=H\sqrt{1-e^2}$. 
Note different range of vertical axis in each panel.
}
\label{fig:fig12}
\end{figure}

In terms of the above instability length scale (\ref{eqn_ins_scale}), 
the renormalized stability diagrams in the ($H^*, \nu$)-plane
are displayed in figures~12($a$), 12($b$), 12($c$) for model $A$, $B$, $C$ and $D$, respectively.
In each panel, we have superimposed three  neutral contours for  
$e=0.9$, $0.99$ and $0.999$ which are indistinguishable from each other
due to the underlying scaling (\ref{eqn_ins_scale1}) with $e$.
In figure~12($b$), we compared the neutral contour of the full model-$B$
with those for model-$B^p$ (that incorporates the EOS for pressure)
and model-$B^\kappa$ (that incorporates  the global EOS for thermal conductivity).
A global equation for thermal conductivity
has little influence on the stability diagram, but a global equation for pressure
significantly enlarges the domain of instability in the ($H^*,\nu$)-plane.

Comparing  figure~12($a$) with  figure~12($b$), we find  that there 
is a significant difference between the predictions
of model-$A$ and model-$B$, especially in the dense limit. 
In particular, with accurate equations of state as in model-$B$, the dense flow is
unstable to shear-banding instability even for small values of the Couette gap.
This is an important issue beyond the square packing density $\nu>\pi/4$ (in two-dimensions)
for which  the flow must reorganize internally such that a part of the flow  forms 
a layered crystalline structure or banding (Alam \& Luding 2003), thereby
allowing the material to  shear.
This reconciles well with the predictions of model-$B$, but not with
model-$A$ (which uses the standard Enskog expressions for all transport coefficients, \S2.1)
for $H^* < 30$ (figure~12$a$); therefore, using an accurate
equation of state over the whole range of densities should give reasonable predictions
for shear-banding solutions.

In this regard, the predictions of model-$C$ (upper curve in figure~12$c$) and 
model-$D$ (lower curve in figure~12$c$) are also
consistent for dense flows since the shear-banding instabilities persist 
at small values of $H$ for these models too.
It may be recalled that these two models ($C$ and $D$),
with  viscosity divergence at $\nu=\nu_\mu<\nu_m$, 
do not admit  ``uniform'' shear solution at large densities $\nu>\nu_\mu$.
However, the related shear-banding solutions at $\nu<\nu_\mu$
can be continued to higher densities $\nu>\nu_\mu$ 
by embedding the present problem into the uniform shear case such
that the shear work is vanishingly small (Khain \& Meerson 2006).

Following one referee's suggestion, we briefly discuss possible
effects of Torquato's (1995) formula for the contact radial distribution function:
\begin{equation}
\begin{array}{rclcl}
    \chi(\nu) &=& \frac{(1 - 0.436\nu)}{(1-\nu)^2},  & \mbox{for} & 0\leq\nu\leq\nu_f, \\
              &=& \frac{(1 - 0.436\nu)}{(1-\nu_f)^2}
                  \frac{(1 - \nu_f/\nu_r)}{(1-\nu/\nu_r)},  & \mbox{for} & \nu_f\leq\nu\leq\nu_r,
\end{array}
\end{equation}
which is known to be valid for an elastic hard-disk fluid over a range of densities up-to
the random packing limit $\nu_r=0.82$.
When (5.6) is used instead of (2.12) in model-$A$, the neutral stability
curve in the ($H^*,\nu$)-plane follows the thick line in figure~12($d$). 
For the sake of comparison, we have also superimposed the neutral stability
curve of model-$A$, denoted by the thin  line, with $\chi(\nu)$ being
given by (2.12) with $\nu_m=\nu_r$.
Clearly, a larger range in the  dense regime is unstable with Torquato's formula (5.6),
but the overall instability characteristics (the stationary nature
of instability, the magnitude of growth rate, etc.) remain similar for both (2.12) and (5.6).
We have also verified that the stability diagram looks similar to that in figure~5 (except for
the presence of a discontinuity on the neutral contour at $\nu=\nu_f$,
resulting in two instability lobes) 
when (2.20) is used as an effective contact radial distribution function
for all transport coefficients.

\subsection{Discussion of some ``ultra-dense'' constitutive models}

Focussing on the ``ultra-dense'' regime with volume fractions close to 
the close-packing density ($\nu\to \nu_m$), the dense limit of model-$A$ 
(i.e. model $A_d$ in \S~2.1.2) can be simplified
by replacing $\nu$ with $\nu_m$ and retaining the dependence of the $f_i$'s on $\nu$
via the corresponding dependence of the pair correlation function.
For this  ultra-dense regime the  constitutive expressions are: 
\begin{equation}
\begin{array}{rcl}
   f_1 \equiv f_1^c(\nu\to\nu_m) &=&   2\nu_m^2\chi\\
   f_2 \equiv f_2^c(\nu\to\nu_m)  &=& 
                 \frac{\sqrt\pi}{8}\left(1 + \frac{8}{\pi}\right)\nu_m^2\chi \\
   f_3 \equiv f_3^c(\nu\to\nu_m) &=& \frac{2}{\sqrt\pi}\nu_m^2\chi\\
   f_4 \equiv f_4^c(\nu\to\nu_m) &=& 
          {\sqrt\pi}\left(\frac{1}{\pi} +\frac{9}{8}\right)\nu_m^2\chi\\ 
   f_5 \equiv f_5(\nu\to\nu_m) &=& \frac{4}{\sqrt{\pi}}(1-e^2)\nu_m^2\chi .
\end{array}
\label{eqn_modelE}
\end{equation}
This model is devoid of  shear-banding instabilities since it can be
verified that
\[
   \frac{{\rm d}\beta_d}{{\rm d}\nu} =0, \quad \forall \quad \nu>0.
\]
This is similar to the predictions of model-$A_d$ (see the dot-dash line in figure~8).

The constitutive model of Khain \& Meerson (2006) can be obtained 
from (\ref{eqn_modelE}) by replacing
the contact radial distribution function for viscosity by
\[
  \chi \to \chi^\mu(\nu) = \left(1 + \frac{0.037}{\nu_\mu -\nu}\right)\chi(\nu)
\]
that diverges at $\nu=\nu_\mu$ as in (2.23); 
the exact density at which viscosity diverges
is not important here. Interestingly, Khain \&  Meerson found two-phase-type
solutions using their model. Since viscosity diverges at a lower density (and hence ``faster'')
than other transport coefficients, it is straightforward to verify that
our instability criterion,
\[
   \frac{{\rm d}\beta_d}{{\rm d}\nu} > 0, \quad \forall \quad \nu>\nu_{cr},
\]
holds for this  model.
Recently, Khain (2007) also  found two-phase-type solutions using 
a constitutive model (a modified model of Khain \& Meerson 2006)
which is similar to our model-$D$,
and the predictions of his model agree well with particle simulation results;
for this modified model too, our instability criterion (\ref{eqn_criterion}) holds.
Therefore, the two-phase-type solutions of Khain \&  Meerson (2006, 2007) are  directly
tied to the shear-banding instabilities of Alam \etal~(1998, 2005, 2006)
via the universal instability criterion (\ref{eqn_criterion}).
More specifically, both belong to the same 
class of {\it constitutive instability} (Alam 2006).

The constitutive model of ~\cite{Losert00} can be obtained 
from (\ref{eqn_modelE}) by using the following functional form for
the contact radial distribution  function 
\[
  \chi  = \left(1 - \nu/\nu_r\right)^{-1},
\]
that diverges at the random close packing limit ($\nu_r$),
for all transport coefficients except the shear viscosity, $\mu$,
that has a ``stronger'' rate of increase near $\nu_r$:
\[
  \chi \to \chi^\mu(\nu) = \left(1 - \nu/\nu_r\right)^{-7/4}.
\]
This choice of viscosity satisfies our instability criterion, ${\rm d}\beta_d/{\rm d}\nu>0$, 
and, therefore, the model of Losert \etal would yield shear-banding type solutions
which is again tied to the increase of ``dynamic'' friction with density
for the uniform shear state.

To illustrate the quantitative effect of a stronger viscosity divergence on the shear-banding
instability, the neutral stability contour for model-$A$ 
(with a stronger viscosity divergence, see below)
is shown in figure 12($d$), denoted by the outermost dotted curve. For this case,
the constititutive model is the same as the full model-$A$ 
(i.e. all transport coefficients diverge at the same density $\nu_m$)
but its viscosity
function $f_2(\nu)$ in (2.8) is calculated using a radial distribution
function that has a  stronger divergence than all other transport coefficients:
\[
   \chi \to \chi^\mu(\nu)  = \frac{1- 7\nu/16}{(1- \nu/\nu_m)^q},
\label{eqn_rdf2}
\] 
with its exponent $q=2.25>2$. 
(It should be noted that this specific functional form of $\chi^\mu$
may not be correct quantitatively at all densities, but it has simply been chosen to
illustrate the possible effects of a stronger viscosity divergence on instabilities.)
Comparing the dotted contour in figure 12($d$) with the thin-solid contour for model-$A$,
we find that a larger range in the ($\nu, H^*$)-plane is unstable for 
the case of a stronger viscosity divergence.
With further increase of $q$, the neutral contour shifts towards the left to cover smaller
values of $H$, leading to even larger instability region in the ($\nu, H^*$)-plane.
In either case, however, the nature of the shear-banding instability remains the same 
and we do not find any new instability as emphasized before.

\subsection{Limit of elastic hard-sphere fluid: dissipation versus effective shear rate}

Naively extrapolating the  instability length-scale  (\ref{eqn_ins_scale1}) to
the elastic limit ($e\to 1$) of atomistic fluids results into $H\to\infty$ that corresponds
to an infinite system for shear-banding to occur in an atomistic fluid.
This is in contrast to molecular dynamics simulations
of sheared ``elastic'' hard-sphere fluid (Erpenbeck 1984)
for which a shear-induced ordering  phenomenon has  been observed
at moderate densities (much below the freezing density).
Similar observations of such banding have been made in simulations for 
continuous potentials too (see, Evans \& Morriss 1986).
The key to resolve this apparent anomaly lies with the fact
that the elastic limit ($e=1$) is singular since the collisional dissipation vanishes.
To achieve a steady-state in simulations of a sheared atomistic fluid, thermostats are used to
take away energy from the system that compensates the production of energy
due to shear-work (${\bf P}:{\bnabla}{\bf u}$). Otherwise
the system would continually heat up, leading to an infinite temperature.
Hence, the collisional dissipation in a granular fluid  can be seen  
to play the role of a thermostat in an atomistic  fluid.

Equating the dissipation term (either due to a thermostat in an elastic fluid, or
due to inelastic collisions in a granular fluid) with the shear-work,
we obtain a scaling relation for temperature with the shear rate and the restitution coefficient:
\begin{equation}
  \tilde{T} \;\;\; \propto\;\; \frac{\gamma^2}{1-e^2}\;\; \propto \;\;   {\gamma^*}^2,
\label{eqn_temp_eff}
\end{equation}
where $\tilde{T}$ is the dimensional temperature, and 
\begin{equation}
  \gamma^* = \frac{\gamma}{\sqrt{1-e^2}}
\label{eqn_srate_eff}
\end{equation}
defined as an ``effective'' shear rate.
For an elastic fluid, this effective shear rate, $\gamma^*$, is used to normalize
the temperature, and hence a similar criterion (\ref{eqn_criterion}) is likely to hold for
the onset of shear-banding instability in an elastic fluid too.
The dependence of the effective  shear rate (\ref{eqn_srate_eff}) with inelasticity
suggests that the shear-banding in atomistic fluids is likely to occur 
at large shear rates, a prediction that agrees with Erpenbeck's (1984) simulations.
This needs to be checked by determining the  analytical expression for the thermostat term
(which might depend on the choice of the thermostat) in the energy equation.

While the Erpenbeck's ordering transition has been explained 
(Kirkpatrik and Nieuwoudt 1986; Lutsko \& Dufty 1986)
as an instability of the ``unbounded'' shear flow of an elastic fluid,
using Navier-Stokes-level equations with wave-vector-dependent transport coefficients
(i.e. generalized hydrodynamics),
the latter work by Lee \etal~(1996) has identified a long-wave instability
(with perturbations along the gradient direction only)
in the uniform shear flow for all densities.
In particular, Lee \etal~ showed that the Navier-Stokes-level constitutive model
is the ``minimal'' model to predict the robustness of this instability. 
The possible connection of this instability with the present work
needs to be investigated in the future.

\subsubsection{Shearbanding criterion in  a molecular fluid: Loose and Hess (1989)}

We close our discussion by recalling a
similar instability criterion for an ``ordering'' transition 
in a dense molecular fluid (Loose and Hess 1989)
and its connection (Alam 2006) with our instability criterion (\ref{eqn_criterion}),
along with a more general criterion for shear-banding in a shear thinning/thickening fluid.

Using a non-Newtonian constitutive model, 
Loose and Hess (1989) have derived a criterion for the
onset of shear-banding in a dense molecular fluid:
\begin{eqnarray}
  \left( \frac{\partial p_{yx}}{\partial {\gamma}} \right)
   \left(\frac{\partial p_{yy}}{\partial\nu} \right) 
& \leq &
   \left(\frac{\partial p_{yx}}{\partial\nu} \right)
  \left( \frac{\partial p_{yy}}{\partial {\gamma}} \right)  ,
\label{eqn_LH1}
\end{eqnarray}
where $p_{yy}$ and $p_{yx}$ are the normal and shear stresses, respectively.
Assuming the following functional dependence of
$p_{yy}$ and $p_{yx}$ with density ($\nu$) and shear rate ($\gamma$),
\begin{equation}
   p_{yy} = p_{yy}^0(\nu) f_{yy}({\gamma}) 
\quad \mbox{and} \quad
   p_{yx} = p_{yx}^0(\nu) f_{yx}({\gamma}) ,
\end{equation}
the above instability criterion simplifies to
\begin{equation}
  \left( p^0_{yx} \frac{{\rm d} f_{yx}}{{\rm d} {\gamma}} \right)
   \left(f_{yy} \frac{{\rm d} p_{yy}^0}{{\rm d}\nu} \right)  \leq
   \left(f_{yx}\frac{{\rm d} p_{yx}^0}{{\rm d}\nu} \right)
  \left( p_{yy}^0 \frac{{\rm d} f_{yy}}{{\rm d} {\gamma}} \right) .
\label{eqn_LH2}
\end{equation}

For a granular fluid, the shear-rate dependence of stresses 
follows the well-known Bagnold scaling:
\begin{equation}
   f_{yy}({\gamma}) \sim T  \sim \gamma^2 
\quad \mbox{and} \quad
   f_{yx}({\gamma}) \sim {\gamma} \sqrt{T} \sim \gamma^2,
\label{eqn_Bagnold1}
\end{equation}
where we have used the relation of the granular temperature with the shear rate,
$T\sim \gamma^2$.
Substituting (\ref{eqn_Bagnold1}) into (\ref{eqn_LH2}),
the Loose-Hess instability criterion boils down to (Alam 2006)
\begin{equation}
  \frac{{\rm d}}{{\rm d}\nu}\left(\frac{p^0_{yx}}{p_{yy}^0}\right) \geq  0 .
\label{eqn_LH3}
\end{equation}
The term within the bracket is the dynamic friction coefficient of a fluid,
and hence the onset of instability is again tied to the increasing value of
this dynamic friction coefficient with increasing density.
This is  same as our shear-banding instability criterion (\ref{eqn_criterion1}).
For a more general case, the shear-rate dependence of stresses
can be postulated as
\begin{equation}
\begin{array}{lcl}
  f_{yy}({\gamma}) &=& {\gamma}^{2n} \\
  f_{yx}({\gamma}) &=& {\gamma}^{n+1},
\end{array}
\end{equation}
where the index $n$ is a measure of shear-thickening ($n>0$)
or shear-thinning ($n<0$) behaviour of the fluid.
With this, the shear-banding instability criterion boils down to
\begin{equation}
  p^0_{yx}\frac{{\rm d} p_{yy}^0}{{\rm d}\nu}
                    \leq
  \left(\frac{2n}{n+1}\right) p_{yy}^0 \frac{{\rm d} p_{yx}^0}{{\rm d}\nu}.
\end{equation}

\section{Conclusion and Outlook}

To conclude, we showed that by just knowing the constitutive expressions
for pressure and shear viscosity, one can determine whether any
Navier-Stokes'-level constitutive model would lead to a shear-banding instability in 
granular plane Couette flow. The onset of this stationary instability
is tied to the increasing value of the ``dynamic'' friction coefficient,
$\beta_d=\mu\gamma/p$ (where $\mu$, $p$ and $\gamma={\rm d}u/{\rm d}y$ 
are the shear viscosity, pressure and shear rate, respectively), 
with increasing density for $\nu>\nu_{cr}$ (equation~(\ref{eqn_criterion})): 
the ``homogeneous'' shear flow breaks 
into alternating layers of dilute and dense regions along the gradient direction
since the flow cannot accommodate the increasing friction to stay
in the homogeneous state.  For $\nu>\nu_{cr}$, the associated ``nonlinear'' 
shear-band solution  corresponds to a lower 
shear stress, or, equivalently, a lower dynamic friction coefficient (Alam 2008).
In other words, the sheared granular flow evolves toward a state of ``lower''
dynamic friction, leading to shear-induced band formation
along the gradient direction.
Note that the  dynamic friction coefficient, $\beta_d=\mu\gamma/p$, is a position-independent
order-parameter for both ``uniform'' ($\gamma=const.$) and ``non-uniform'' ($\gamma=\gamma(y)$)
shear flows.

In the framework of a ``dense'' constitutive model that incorporates only collisional
transport mechanism (i.e. Haff's model, 1983), 
we showed that an accurate global equation of state for
pressure or a viscosity divergence at a lower density
(with other transport coefficients being given by respective
Enskog values) can induce  shear-banding instabilities, even though
the original dense Enskog model is stable to such shear-banding instabilities.
Since the prediction of the shear-banding instability depends crucially on the form of
the constitutive relations, we need to use  accurate forms
of constitutive expressions over the whole range of density
that incorporate both kinetic and collisional transport mechanisms. 
The latter statement is important 
since the dilute and dense regimes coexist even at a low mean density
when the uniform shear flow is unstable to shear-banding instability.
The resulting nonlinear shear-banding solutions of all four models ($A$, $B$, $C$ and $D$)
and their stability as well as the related results from 
particle-dynamics simulations will be considered in a separate paper.
This will help to identify the best among these four constitutive models,
or will point towards new models.
In this regard, it is recommended that the particle-dynamics simulations
be used to find out accurate expressions 
(valid over the whole range of density) for all transport coefficients.

We established  that the two-phase-type solutions of 
Khain \&  Meerson (2006) are  directly related 
to the shear-banding instabilities of Alam \etal~(1998, 2005, 2006)
via the universal instability criterion (\ref{eqn_criterion}),
and both belong to the same class of {\it constitutive instability} (Alam 2006).
In particular, the instabilities arising out of non-monotonicities of 
constitutive relations with mean-fields (\eg the coil-stretch transition
is tied to non-monotonic stress-strain curve; see, de Gennes 1974) are 
of constitutive origin and hence dubbed constitutive instability. 
The same universal criterion (\ref{eqn_criterion}) also
holds for the constitutive model of Losert \etal~(2000),
thereby yielding  such two-phase-type solutions in 
their model of plane shear flow.

The onset of the ordering  transition of Erpenbeck (1984) in
a sheared ``elastic'' hard-sphere fluid (which is close to our granular system)
is  accompanied by a decrease in viscosity and hence a lower viscous dissipation.
Therefore, similar to the sheared granular fluid, the state of lower viscosity/friction
is the preferred equilibrium state for a sheared  atomistic fluid.
Our  instability criterion (\ref{eqn_criterion})
seems to provide a unified description for the shear-banding phenomena 
for the singular case of hard-sphere fluids 
if we relate the collisional dissipation
to a thermostat, leading to an ``effective'' shear rate.
This possible connection needs to be investigated further from
the viewpoint of a constitutive instability of the underlying field equations.

The shear-banding phenomenon is ubiquitous in a variety of complex fluids under
non-equilibrium conditions: wormlike micelles (Spenley, Cates \& McLeish 1993), 
colloidal suspensions (Hoffman 1972; Ackerson \& Clark 1984),
model glassy material (Varnik \etal~2003),
suspensions of rod-like viruses (Lettinga \& Dhont 2004),
lyotropic liquid crystals (Olmsted 2008) and numerous other systems. 
In the literature of non-Newtonian fluids (see, for a review,
Olmsted 2008), the shear-banding phenomenon
has been explained as a constitutive instability
from the linear stability analysis of appropriate
constitutive models (Greco \& Ball 1997; Wilson \& Fielding 2005).
The well-known ``Hoffman-transition'' in a colloidal suspension
(banding/ordering of particles along the gradient direction) above the freezing density
is accompanied by a sharp decrease in viscosity and has been
explained in terms of a flow-instability (Hoffman 1972). 
A very recent work (Caserta, Simeone \& Guido 2008) on biphasic liquid-liquid
systems  showed that the shear-induced banding in such systems
is tied to a lower viscosity, or, equivalently, a lower viscous dissipation.
For both cases, the criterion of lower viscosity is similar to 
our criterion of a lower ``dynamic'' friction
for the band-state in sheared granular fluid.
It appears that the shear-induced banding in many 
complex fluids has a common theoretical description in terms of
``constitutive'' instability that leads to an ordered-state of
a lower viscosity/friction.




\begin{thebibliography}{99}
\bibitem[Ackerson \& Clark  (1984)]{AC84}
\textsc{Ackerson, B. J. \& Clark, N. A.} 1984
Shear-induced partial translational ordering of a colloidal solid. 
\textit{Phys. Rev. A} {\bf 30}, 906.
\bibitem[Alam (2005)]{Alam05a}
\textsc{Alam, M.} 2005
Universal unfolding of pitchfork bifurcations and the shear-band
formation in rapid granular Couette flow.
In \textit{Trends in Applications of Mathematics to Mechanics}
(ed. Y. Wang \& K. Hutter), pp. 11-20, Shaker-Verlag, Aachen.
\bibitem[Alam (2006)]{Alam06}
\textsc{Alam, M.} 2006
Streamwise structures and density patterns in rapid granular Couette flow: 
a linear stability analysis.
\textit{J. Fluid Mech.} {\textbf 553}, 1.
\bibitem[Alam (2008)]{Alam08}
\textsc{Alam, M.} 2008
Dynamics of sheared granular fluid.
In \textit{Proc. of 12th  Asian Congress of Fluid Mechanics}
(ed. H.J. Sung), pp. 1-6,
18--21 August, Daejeon, Korea.
\bibitem[Alam \etal~(2005)]{Alam05}
\textsc{Alam, M., Arakeri, V., Goddard, D., Nott, P. \& Herrmann, H.} 2005
Instability-induced ordering, universal unfolding and the role of gravity
in granular Couette flow.
\textit{J. Fluid Mech.} {\textbf 523}, 277.
\bibitem[Alam \& Luding (2003a)]{AL03a}
\textsc{Alam, M. \&  Luding, S.} 2003a
Rheology of bidisperse granular mixture via event-driven simulations.
\textit{J. Fluid Mech.} {\textbf 476}, 69.
\bibitem[Alam \& Luding (2003)]{AL03}
\textsc{Alam, M. \&  Luding, S.} 2003
First normal stress difference and crystallization in a dense sheared granular fluid.
\textit{Phys. Fluids} {\textbf 15}, 2298.
\bibitem[Alam \& Luding (2005)]{AL05}
\textsc{Alam, M. \&  Luding, S.} 2005
Energy noequipartition, rheology and microstructure in sheared bidisperse granular mixtures.
\textit{Phys. Fluids} {\textbf 17}, 063303.
\bibitem[Alam \& Nott (1998)]{AN98}
\textsc{Alam, M. \&  Nott, P. R.} 1998
Stability of plane Couette flow of a granular material.
\textit{J. Fluid Mech.} {\textbf 377}, 99.
\bibitem[Alam \& Nott (1997)]{AN97}
\textsc{Alam, M. \&  Nott, P. R.} 1997
Influence of friction on the stability of unbounded granular shear flow.
\textit{J. Fluid Mech.} {\textbf 343}, 267.
\bibitem[Campbell (1990)]{Campbell90}
\textsc{Campbell, C. S.} 1990  
Rapid granular flows. 
\textit{Annu. Rev.  Fluid Mech.} {\textbf 22}, 57.
\bibitem[Caserta, Simeon \& Guido (2008)]{CSG90}
\textsc{Caserta, S., Simeon, M. \& Guido, S.} 2008
Shearbanding in biphasic liquid-liquid systems.
\textit{Phys. Rev. Lett.} {\bf 100}, 137801.
\bibitem[Conway \& Glasser (2004)]{CG04}
\textsc{Conway, S. \& Glasser, B. J.} 2004
Density waves and coherent structures in granular Couette flow.
\textit{Phys. Fluids} {\bf 16}, 509.
\bibitem[de Gennes (1974)]{deGennes74}
\textsc{de Gennes, P. G.} 1974
Coil-stretch transition of dilute flexible polymers under ultra-high gradients.
\textit{J. Chem. Phys.} \textbf{60}, 5030.
\bibitem[Erpenbeck (1984)]{Erpenbeck84} 
\textsc{Erpenbeck, J.} 1984 
Shear viscosity of the hard sphere fluid via non-equilibrium molecular dynamics.
\textit{Phys. Rev. Lett.} {\bf 52}, 1333.
\bibitem[Evans \& Morriss (1986)]{EM86}
\textsc{Evans, D. J. \&  Morriss, G. P.} 1986 
Shear thickening and turbulence in simple fluids.
\textit{Phys. Rev. Lett.} {\bf 56}, 2172.
\bibitem[Garcia-Rojo \etal~(2006)]{GLB06}
\textsc{Garcia-Rojo, R.,  Luding, S. \&   Brey, J. J.} 2006  
Transport coefficients for dense hard-disk systems.
\textit{Phys. Rev. E} {\bf 74}, 061305.
\bibitem[Gass (1971)]{Gass71}
\textsc{Gass, D. M.} 1971  
Enskog theory for a rigid disk fluid.
\textit{J. Chem. Phys.} {\bf 54}, 1898.
\bibitem[Gayen \& Alam (2006)]{GA06}
\textsc{Gayen, B. \& Alam, M.} 2006
Algebraic and exponential instabilities in a sheared micropolar granular fluid.
\textit{J. Fluid Mech.} {\bf 567}, 195.
\bibitem[Gayen \& Alam (2008)]{GA08}
\textsc{Gayen, B. \& Alam, M.} 2008
Orientational correlation and velocity distributions in uniform shear flow of a dilute granular gas.
\textit{Phys. Rev. Lett.} {\bf 100}, 068002.
\bibitem[Greco \& Ball (1997)]{GB97}
\textsc{Greco, F. \& Ball, R. C.} 1997
Shear-band formation in a non-Newtonian fluid model with a constitutive instability.
\textit{J. Non-Newt. Fluid Mech.} {\bf 69}, 195.
\bibitem[Goldhirsch (2003)]{Goldhirsch03}
\textsc{Goldhirsch, I.} 2003
Rapid granular flows.
\textit{Annu. Rev. Fluid Mech.} {\textbf 35}, 267.
\bibitem[Haff (1983)]{Haff83}
\textsc{Haff, P. K.} 1983
 Grain flow as a fluid-mechanical phenomenon.
\textit{J. Fluid Mech.} {\bf 134}, 401.
\bibitem[Henderson (1975)]{Henderson75}
\textsc{Henderson, D.} 1975 
A simple equation of state for hard discs.
\textit{Mol. Phys.} {\bf 30}, 971.
\bibitem[Hopkins \& Louge (1991)]{HL91}
\textsc{Hopkins, M. \& Louge, M. Y.} 1991
Inelastic microstructure in rapid granular flows of smooth disks.
\textit{Phys. Fluids A} {\bf 3}, 47.
\bibitem[Hoffman (1972)]{Hoffman72}
\textsc{Hoffman, R. L.} 1972
Discontinuous and dilatant viscosity behaviour in concentrated suspensions.
I. Observation of a flow instability. 
\textit{Trans. Soc. Rheol.}, {\bf 16}, 155.
\bibitem[Jenkins \& Richman (1985)]{JR85}
\textsc{Jenkins, J. T. \& Richman, M. W.} 1985
Kinetic theory for plane flows of a dense gas of identical, rough, inelastic, circular disks.
\textit{Phys. Fluids} {\bf 28}, 3485.
\bibitem[Kirkpatrick \& Nieuwoudt (1986)]{KN86}
\textsc{Kirkpatrick, T. R. \& Nieuwoudt, J. C.} 1986
Stability analysis of a dense hard-sphere fluid subjected to large shear-induced ordering.
\textit{Phys. Rev. Lett.} {\bf 56}, 885.
\bibitem[Khain  (2007)]{Khain07}
\textsc{Khain, E.} 2007
Hydrodynamics of fluid-solid coexistence in dense shear granular flow.
\textit{Phys. Rev. E} {\bf 75}, 051310.
\bibitem[Khain \& Meerson (2006)]{KM06}
\textsc{Khain, E. \&  Meerson, B.} 2006
Shear-induced crystallization of a dense rapid granular flow: Hydrodynamics
beyond the melting point.
\textit{Phys. Rev. E} {\bf 73}, 061301.
\bibitem[Lee \etal (1996)]{LDMSL96} 
\textsc{Lee, M.,  Dufty, J. W., Montanero, J. M., Santos, A. \& Lutsko, J. F.} 1996 
Long wavelength  instability for uniform shear flow.
\textit{Phys. Rev. Lett.} {\bf 76}, 2702.
\bibitem[Lettinga \& Dhont (2004)]{KD04}
\textsc{Lettinga, M. P. \& Dhont, J. K. G.} 2004
Non-equilibrium phase behaviour of rodlike viruses under shear flow.
\textit{J. Phys. Cond. Matter} {\bf 16}, S3929.
\bibitem[Loose \&  Hess (1989)]{LH89}
\textsc{Loose, W. \&  Hess, S.} 1989
Rheology of dense model fluids via nonequilibrium molecular dynamics:  
shear-thinning and ordering transition.
\textit{Rheol. Acta} {\bf 28}, 101.
\bibitem[Losert \etal~(2000)]{Losert00}
\textsc{Losert, W., Bocquet, L.,  Lubensky, T. C. \& Gollub, J. P.} 2000
Particle dynamics in sheared granular matter.
\textit{Phys. Rev. Lett.} {\bf 85}, 1428.
\bibitem[Luding (2001)]{Luding01}
\textsc{Luding, S.} 2001
Global equation of state of two-dimensional hard-sphere systems. 
\textit{Phys. Rev. E} {\bf 63}, 042201.
\bibitem[Luding (2002)]{Luding02}
\textsc{Luding, S.} 2002 
Liquid-solid transition in bidisperse granulates.
\textit{Adv. Complex Syst.} {\bf 4}, 379.
\bibitem[Luding (2008)]{Luding08}
\textsc{Luding, S.} 2008
From dilute to very dense granular media.
\textit{In preparation}.
\bibitem[Lun \etal~(1984)]{LSJC84}
\textsc{Lun, C. K. K., Savage, S. B.,  Jeffrey, D. J. \& Chepurniy, N.} 1984
Kinetic theories for granular flow: inelastic particles in Couette flow and slightly
inelastic particles in a general flow field.
\textit{J. Fluid Mech.} {\bf 140}, 223.
\bibitem[Lutsko \& Dufty (1986)]{LD86} 
\textsc{Lutsko, J. F. \& Dufty, J. W.} 1986 
Possible instability for shear-induced order-disorder transition.
\textit{Phys. Rev. Lett.} {\bf 57}, 2775.
\bibitem[McNamara (1993)]{TG93}
\textsc{McNamara, S.} 1993
Hydrodynamic modes of a uniform granular medium.
\textit{Phys. Fluids A} {\bf 5}, 3056.
\bibitem[Mueth {et al.} (2000)]{Mueth00}
\textsc{Mueth, D. M., Debregeas, G. F., Karczmar, G. S., Eng, P. J., Nagel, S. \& Jaeger, H. J.} 2000
Signatures of granular microstructure in dense shear flows.
\textit{Nature} {\bf 406}, 385.
\bibitem[Nott \etal~(1999)]{NAAJS99}
\textsc{Nott, P. R., Alam, M., Agrawal, K., Jackson, R. \& Sundaresan, S.} 1999
The effect of boundaries on the plane Couette flow of granular materials: a bifurcation analysis.
\textit{J. Fluid Mech.} {\textbf 397}, 203.
\bibitem[Olmsted (2008)]{Olmsted08} 
\textsc{Olmsted, P. D.} 2008
Perspective on shear banding in complex fluids.
\textit{Rheol. Acta} {\bf 47}, 283.
\bibitem[Saitoh \& Hayakawa (2007)]{SH07}
\textsc{Saitoh, K. \& Hayakawa, H.} 2007
Rheology of a granular gas under a plane shear.
\textit{Phys. Rev. E} {\bf 75}, 021302.
\bibitem[Savage \&  Sayed (1984)]{SJ84}
\textsc{Savage, S. B. \& Sayed, S.} 1984
Stresses developed by dry cohesionless granular materials sheared
in an annular shear cell.
\textit{J. Fluid Mech.} {\bf 142}, 391.
\bibitem[Sela \& Goldhirsch (1998)]{SG98}
\textsc{Sela, N. \& Goldhirsch, I.} 1998
Hydrodynamic equations for rapid shear flows of smooth, inelastic
spheres, to Burnett order.
\textit{J. Fluid Mech.} {\bf 361}, 41.
\bibitem[Shukla \& Alam (2008)]{SA08}
\textsc{Shukla, P. \& Alam, M.} 2008
Nonlinear stability of granular shear flow: Landau equation, shearbanding and universality.
In \textit{Proc. of International Conference on Theoretical and Applied Mechanics}
(ISBN 978-0-9805142-0-9),
24--29 August, Adelaide, Australia.
\bibitem[Spenley, Cates \& McLeish (1993)]{SCM93}
\textsc{Spenley, N. A., Cates, M. E. \& McLeish, T. C. B.} 1993
Nonlinear rheology of wormlike micelles.
\textit{Phys. Rev. Lett.} {\bf 71}, 939.
\bibitem[Tan \& Goldhirsch (1997)]{TG97}
\textsc{Tan, M. \& Goldhirsch, I.} 1997
Intercluster interactions  in rapid granular shear flows.
\textit{Phys. Fluids} {\bf 9}, 856.
\bibitem[Torquato (1995)]{Troquato95}
\textsc{Torquato, S.} 1995
Nearest-neighbour statistics for packings of hard spheres and disks.
\textit{Phys. Rev. E} {\bf 51}, 3170.
\bibitem[Tsai \etal~(2003)]{Gollub03}
\textsc{Tsai, J. C., Voth, G. A.  \& Gollub, J. P.} 2003
Internal Granular dynamics, shear-induced crystallization, and compaction steps.
\textit{Phys. Rev. Lett.} {\bf 91}, 064301.
\bibitem[Varnik \etal~(2003)]{VBBB3}
\textsc{Varnik, F., Bocquet, L., Barrat, J.-L. \& Berthier, L.} 2003
Shear localization in model glass.
\textit{Phys. Rev. Lett.} {\bf 90}, 095702.
\bibitem[Volfson \etal~(2003)]{VTA03}
\textsc{Volfson, D., Tsimring L. S. \&  Aranson, I. S.} 2003
Partially fluidized shear granular flows: Continuum theory and molecular dynamics simulations.
\textit{Phys. Rev. E} {\bf 68}, 021301.
\bibitem[Wilson \& Fielding (2005)]{WF05}
\textsc{Wilson,  H. J. \& Fielding, S. M.} 2005
Linear instability of planer shear banded flow of both diffusive and non-diffusive
Johnson-Segelman fluids.
\textit{J. Non-Newt. Fluid Mech.} {\bf 138}, 181.
\end{thebibliography}
\end{document}